\documentclass[12pt]{article}
\usepackage{latexsym}
\usepackage{amssymb}
\usepackage{amsmath}
\usepackage{graphicx}
\usepackage{enumitem}
\usepackage{slashed}
\usepackage{hyperref}

\newcommand\eeq{\end{equation}} 
\newcommand\beq{\begin{equation}}

\def\be{\begin{equation}}
\def\ee{\end{equation}}
\def\bear{\begin{eqnarray}}
\def\eear{\end{eqnarray}}

\newcommand\bra[1]{{\langle {#1}|}}
\newcommand\ket[1]{{|{#1}\rangle}}

\def\a{\alpha}

\def\bra{\langle}
\def\ket{\rangle}
\def\a{\alpha}

\newcommand{\ti}[1]{\tilde{#1}}

\newcommand{\sm}[1]{\mbox{\scriptsize #1}}

\renewcommand{\@}[1]{\sqrt{#1}}
\renewcommand{\le}[1]{\label{#1}\end{eqnarray}}
\newcommand{\bea}{\begin{eqnarray}}
\newcommand{\eea}{\end{eqnarray}}

\newcommand{\eq}[1]{(\ref{#1})}

\def\ffract#1#2{\raise .35 em\hbox{$\scriptstyle#1$}\kern-.25em/
\kern-.2em\lower .22 em \hbox{$\scriptstyle#2$}}

\begin{document}

\pagestyle{empty}

\centerline{{\Large \bf  On Symmetry and Duality}}
\vskip 1.5truecm

\begin{center}
{\large Sebastian De Haro$^{1,2,3}$ and Jeremy Butterfield$^1$}\\
\vskip 1truecm
$^1${\it Trinity College, Cambridge, CB2 1TQ, United Kingdom}\footnote{Published online in {\it Synthese,} in a special issue on `Symmetries and Asymmetries in Physics' edited by M.~Frisch,  R.~Dardashti, and G.~Valente.}\\
$^2${\it Department of History and Philosophy of Science, University of Cambridge}\\
$^3${\it Vossius Center for History of Humanities and Sciences, University of Amsterdam}

\vskip .7truecm
{\tt sd696@cam.ac.uk, jb56@cam.ac.uk}
\vskip 1cm
\today
\end{center}

\vskip 6truecm

\begin{center}
\textbf{\large \bf Abstract}
\end{center}

We advocate an account of dualities between physical theories: the basic idea is that dual theories are isomorphic representations of a common core. We defend and illustrate this account, which we call a Schema, in relation to  symmetries. 

Overall, the account meshes well with standard treatments
of symmetries. But the distinction between   the common core and the dual theories prompts a distinction between three kinds of symmetry: which we call `stipulated', `accidental' and `proper'.

\newpage
\pagestyle{plain}

\tableofcontents

\newpage

\section{Introduction}\label{intro}

Symmetries have  long been a central topic in the philosophy of physics: as witness collections such as Brading and Castellani (2003) and this special issue. But dualities between physical theories have only more recently become a focus of interest.\footnote{See for example Castellani (2017), Dieks et al.~(2015), Fraser (2017), Huggett (2017), Matsubara (2013), Read (2016), Rickles (2017).\label{refs}} In this paper we bring the topics together, by applying an account of dualities that we have developed elsewhere.

Bringing the topics together is natural.  For in the literature, it is agreed by all hands that a duality is like a ``giant symmetry'': a symmetry between theories. For in physics, the basic idea of a symmetry is a map taking a  state of the system into another appropriately related state; and correspondingly mapping physical quantities---details below. And in a duality, an entire theory is mapped into another appropriately related theory.

Our account of dualities will confirm this basic analogy.  The leading idea will be: (i) the preservation, or appropriate matching, of a state's values for various quantities, and (ii) this preservation or matching being maintained by the dynamics of the system. (We say `preservation or matching' so as to respect the distinction between invariance and covariance, and `dynamics' is to include Euclidean systems: details below.)

In short: our account---we call it a `Schema'---holds that a duality between two theories  requires that:\\
\indent \indent (a): the two theories share a {\em common core}; the common core is itself a theory, which we will call the {\em bare theory}; 
(for us, a {\it bare} theory or model is a theory or model stripped of its interpretation; more details in Section \ref{thymodel}); and \\
\indent \indent  (b): the two given theories are {\em isomorphic models} of this common core, the bare theory.\\
Here, we understand `models' as realizations or formulations. They are almost always representations in the sense of representation theory, i.e.~homomorphic copies of the bare theory. Representations are of course in general {\em not} isomorphic. But we say that duality is a matter of two models of a common core, a bare theory, being indeed isomorphic with respect to the structure of that core. 

We stress  that here we mean `representation' in the mathematical, not the philosophical, sense. The  bare dual theories do not interpret the bare common core theory (notwithstanding our use of the word `model'). Rather they are specific realizations or formulations of the bare theory, like the matrix representations of an abstract group. But agreed: the dual theories, and also often  the bare theory, {\em do} get interpreted. Interpretation, and our choice of jargon, is further explained in Section \ref{thymodel}. 

We have developed this Schema in various papers. The most detailed one is our (2018).  It also illustrates the Schema with a major example: bosonization, which is a duality between two quantum field theories in two spacetime dimensions---one with bosons, and one with fermions. So here, a bosonic model  is isomorphic to a fermionic model; (their common core is a certain infinite-dimensional algebra). Other papers (De Haro (2016, 2018a), Butterfield (2018))  discuss in more detail duality's relation to the interpretation of theories, especially  the various senses in which theories get called `equivalent'. Other papers discuss other advanced, indeed conjectural, examples in string theory (De Haro (2015, 2017), De Haro et al.~(2016, 2017)). And yet  other papers discuss the less formal aspects of dualities: in particular,  their heuristic role in finding a theory ``behind'' the bare theory, of which the dual models are only approximations not representations (De Haro (2018)), and in relations to understanding (De Haro and De Regt (2018, 2018a)).\footnote{We admit at the outset that because our Schema defines duality formally, without regard to the isomorphism being surprising or scientifically important, it has many illustrations that are unremarkable, and indeed not usually called dualities. We will see examples below: cf.~Section \ref{explematriGalil}. Contrast a case like bosonization. Since bosons and fermions are very different, it is indeed a surprising isomorphism.  It is also scientifically important because it pairs situations of strong coupling in one dual (so that problems are in general difficult to solve, since perturbation theory cannot be trusted) with situations of weak coupling in the other dual: so one can sometimes solve a problem in the weak coupling regime of one dual and transfer the result, so as to solve an intractable problem in the other dual. Broadly speaking, dualities' scientific importance depends on these features: the isomorphism being surprising, and its relating strong and weak coupling regimes. But features like importance and surprise are hard to be precise about: hence our decision that it is best to define duality formally, even at the expense of countless trivial examples.  For more discussion, cf.~De Haro and Butterfield (2018:~Section 2.1) and De Haro (2018:~Sections 1, 3.3).\label{logiweak}}

Thus the aim of this paper is to relate the Schema to symmetries in more detail than we have done previously. (Our  previous discussion was in (2018: Sections 3.1.1, 3.2.4).) The main topic will be the relations between symmetries of the common core theory, and symmetries of one of the dual theories (one of the models of the common core). This topic is an important preliminary to discussing the `interaction' of symmetries with dualities. It will call for care about what parts of a model ``do the representing'' of a common core; (cf. Section \ref{rootspec}'s distinction between {\em model root} and {\em specific structure}). It will also lead to a classification of kinds of symmetry (Section \ref{symmths}, summarized in Section \ref{classify}). Thus we will give as much emphasis to symmetries as to dualities (usually treating symmetries first). 
 
To be as clear as possible, we will build up the details successively. In effect, Sections \ref{jargonnot} and \ref{schema} set the stage for the main claims in the second half of the paper. We begin by explaining the contrast between the two `levels': the bare theory, and its models (realizations). This contrast is a prerequisite to our discussing symmetries and dualities, since it can  (should!) be explained for the case of a single model (realization): i.e.~regardless of whatever dualities, or symmetries, may hold good. This we do in Section \ref{jargonnot}. As we will see, this involves various issues of interpretation, even controversy. And it prompts our distinction between {\em model root} and {\em specific structure}.

We are thereby equipped to give a brief exposition of our Schema, and a more detailed prospectus of the second half of the paper---i.e.~of our account of the relations between symmetries and dualities. These are in Section \ref{schema}. 

This account starts with symmetries in general: without regard to dualities, or even our distinction between theories and the models that realize or represent them; (Section \ref{symmygenl}). Here, we relate our basic idea of symmetry,  as a map on states that preserves quantities' values, to some familiar topics: such as dynamical symmetries, how spacetime theories treat symmetries, and what is the subset of quantities that it is ``worthwhile'' for symmetries to preserve (cf. Caulton 2015). Then we discuss how symmetries fit with the theory-model relation: but again without regard to dualities; (Section \ref{symmths}). Here, the differing amounts of structure at the levels of the bare theory and its models---the distinction between  model root and specific structure---will prompt definitions of three kinds of symmetry: which we call {\em stipulated}, {\em accidental} and {\em proper}. Finally in Section \ref{reltosymmy}, we describe how on our account, dualities preserve symmetries in a straightforward way. Section \ref{conclude} concludes.

A final preliminary: our scope is limited. We will not engage in detail with two debates that have been prominent in the recent philosophical literature about symmetries. For these debates are orthogonal to most of this paper's issues: which centre around the relations between symmetries of the common core, and symmetries of its models. But for completeness, we here briefly note these debates, and our broad view about them. They both concern whether symmetry is always a sign of `surplus structure', `redundancy' or `gauge': for the relation of this to duality, cf. our (2018: Section 3.2.4)  \\
\indent \indent (1): Should two symmetry-related solutions of a theory: (a) be interpreted {\em ab initio} as representing the same physical state of affairs?; or (b) be taken merely to motivate searching for a common ontology that secures such an interpretation? This debate is articulated by M\o ller-Nielsen (2017) and, in relation to dualities, Read and M\o ller-Nielsen (2018). These authors defend option (b). Broadly speaking, we agree with them, about symmetries as well as dualities: (cf. De Haro (2016:~Section 1.3), Butterfield (2018:~Sections 1.2, 3.3) and De Haro (2018a:~Section 2.3). \\
\indent \indent (2): Given a theory whose solution-space is partitioned by a group of symmetries---i.e.~solutions in the same same cell are symmetry-related---should we: (a) try to write down a `quotiented' (also known as: `reduced') theory whose solutions correspond to the cells of the partition; or (b) resist quotienting the given theory, but take its symmetry-related solutions to be isomorphic? This debate is articulated, with (a) and (b) labelled `reduction' and `sophistication' respectively, by Dewar (2015:~Sections 4 to 6; 2017); see also Caulton (2015:~pp.~156-157). Dewar defends sophistication. Broadly speaking, we are sympathetic (De Haro (2018a:~Section 3.2.3) and Butterfield (2018: Section 2.3)): reduction is not to be undertaken lightly. But the Schema does not commend one or the other: we will encounter these two alternatives in Section \ref{expleGalil}, and we will see that the choice between a `reduced' or a `sophisticated' formalism basically comes down to a choice of representations. 

\section{Theories, Models, Model Roots and Specific Structure}\label{jargonnot}

We said at the beginning of Section \ref{intro} that on our account of duality,  two dual theories share a common core, which is itself a theory, the bare theory---and that they are isomorphic models of it. So to explain that account, we first need to explain our use of the terms `theory' and `model': which, after all, have various (conflicting!) uses. We do this in Section \ref{thymodel}. This leads to: distinguishing within a model, what we will call the {\em model root} (in most cases: a {\em model triple}) from the {\em specific structure} (Section \ref{rootspec}); and giving examples of this distinction (Section \ref{explematriGalil}). Finally, we introduce notation for the values of physical quantities on states (Section \ref{values}). Then we will be ready to define duality as an isomorphism of model roots: details in Section \ref{schema}. 

\subsection{`Theory' vs.~`model': interpretation}\label{thymodel}

For the words `theory'  and `model', the first point to make---the main idiosyncrasy in our usage---arises from the fact that since a duality compares two theories, it involves two ``levels'': the common core ``above'' (more general and-or more abstract) and the theories ``below''  (more specific and-or more concrete). On the other hand,  one naturally thinks of a theory as general and-or abstract, ``standing above'', and ``being common to'', its more specific and-or concrete models. So in order to discuss duality, the question arises: should we allocate the word `theory' to the common core, or to each of the two dual theories? 

We opt for the former. And since the two dual theories (so-called!) are indeed specific and-or concrete  realizations (versions, formulations) of this common core that we have opted to call a `theory',  we also co-opt the word `model' for the duals themselves. So the overall effect of our jargon is to lift the use of the words `theory' and `model', ``one level up".

As a simple example to illustrate our usage, consider position-momentum duality in elementary Schr\"odinger-picture quantum mechanics. Although the position and momentum representations might have been discovered, in a counterfactual history, as different ``theories'' (since the Schr\"odinger equation takes very different forms in the position and momentum bases, as do the operators and wave-functions), the discovery of the Fourier transformation reveals that these are in fact two representations of a common core theory: namely, quantum mechanics formulated in the basis-free language of Hilbert space. Thus what in the counterfactual history is thought to be two distinct ``theories'', is seen to be just two models, i.e.~representations, of a common structure, which we call elementary quantum mechanics.

\subsubsection{`Model'}\label{model}

So beware! This means that our use of `model' rejects three common connotations of the word.\footnote{In Section \ref{fail}, we will recover the first notion, (i), of a `model', as a {\it special case} of our usage. That we here reject this connotation means that our models are not {\em defined} as in (i). But in Section \ref{fail}'s specific case, ``models'' in sense (i) turn out to be models in our sense.\label{usualusage}} Namely, the connotations that a model is: \\
\indent (i) a solution of the theory, or a history of the system concerned (often a trajectory through the state-space), as against `all solutions'---which are the purview of the theory; and-or \\
\indent (ii) an approximation(s) to what the theory says (maybe about a specific regime or system the theory is concerned with); and-or \\
\indent (iii) a part of the empirical world---a hunk of reality!---that thus gives an interpretation (or part of an interpretation) of the theory. \\
We will indeed later be concerned with the ideas (i) to (iii): (in fact, with (iii) in this Subsection). But they are not part of our definition of `model'.  

Instead, a model is for us a specific realization (or version or formulation) of a theory. That is: it `models' (verb!) another theory ``above'', which in general also has other such models.  Almost always, it is a representation of the theory above, in the sense of mathematical  representation theory---`representation' being another word with confusingly diverse uses. Recall the example, in this Section's preamble, of position vs.~momentum representations.

And as we have announced: the main case for us of ``the theory above''  will be the {\em common core}, the {\em bare theory}, of two dual theories (our `models'). (It is usually best to think of the bare theory as uninterpreted, or abstract: though it may be interpreted: cf.~Section \ref{interp}). So we say the two given theories are {\em models} of the bare theory; and their being {\em isomorphic} models of the bare theory (isomorphic as regards the structure of the bare theory) is what makes them duals. 

\subsubsection{`Theory'}\label{theory}

As to our use of `theory', the main thing to say is that it is mainstream: i.e.~typical of the literature, especially the literature on the semantic conception of theories as applied to physics. Agreed: recent philosophy of science has emphasised many aspects of scientific endeavour that hardly invoke the notion of {\em scientific theory}, central though this notion was for discussions by both the positivists and their successors. For example, aspects such as experiment (calibration of instruments etc.), causation (mechanistic explanation etc.), and the social dimensions of knowledge (testimony etc.) have recently been discussed with a strong emphasis on models (in a more usual sense than ours!), rather than theories.  

We agree that these aspects of scientific endeavour are important for our philosophical understanding of science (De Haro and De Regt (2018: Section 1.1)). But even if these aspects do not need the notion of theory, still the notion may well be useful for other aspects. Indeed, we believe it is indispensable for discussion of symmetries and dualities in physics.\footnote{We also believe it useful, even indispensable, in other discussions. One main one is understanding renormalization---a topic for which, again, there has been scepticism about its usefulness: e.g.~Kaiser (2005:~pp.~377-387). For a defence of the notion, cf.~Butterfield (2014:~Section IV.1).}

More specifically: we can often think of a physical theory as a triple: a state-space $\cal S$ (e.g.~a Hilbert space in a quantum theory), a set of physical quantities $\cal Q$ (almost always an algebra), and a dynamics $\cal D$ that describes how the values of quantities (on states) change over time (and-or over space---so as to accommodate Euclidean theories). As we said in Section \ref{intro}, symmetries and dualities will concern the preservation or matching of these values.\footnote{In addition, a theory often comes with stipulated symmetries. We will discuss these in Sections \ref{salientstipul} and \ref{secure}.\label{Twsyms}}  So we will sometimes write a theory (and of course: a  model in our sense) as a triple, e.g.~$\bra {\cal {S}}, {\cal {Q}}, {\cal {D}} \ket$. In any specific case, each of $\cal S$, $\cal Q$ and $\cal D$ will of course have further structure. For example:\\
\indent (i):  for a quantum theory, $\cal S$ will usually be a Hilbert space, or a set of density matrices; for a classical theory, it will usually  be a manifold, or a set of probability distributions; \\
\indent (ii):  for almost any theory, $\cal Q$ will be an algebra over the real or complex numbers, allowing quantities to be added and multiplied; and \\
\indent (iii): for almost any theory, the dynamics $\cal D$ can be understood either in Schr\"{o}dinger picture, with states changing over time and quantities fixed over time, or in Heisenberg picture (vice versa). 

Two clarifications about our treatment of a theory as a triple: the first formal, the second interpretative. First: note that we said: `we can {\em often} think of a physical theory as a triple', and `{\em almost any}' in (ii) and (iii). For we agree that not every theory is presented, or best thought of, in this way. Theories in statistical and quantum physics are often formulated in terms of partition functions and-or path integrals with sources, and related concepts like sets of correlation functions, rather than in terms of states and quantities. And in field theories, the dynamics is often presented as field equations holding at each spacetime point---and so not naturally thought of in terms of the Schr\"{o}dinger or Heisenberg pictures with their ``background time''. But in this paper, we can think of theories (and models, in our sense) as such triples: all our morals will carry over to these other ways of formulating theories.\footnote{Of course, one can draw connections between formulations with states, quantities and dynamics---our triple conception---and other  types of formulation, like  partition functions, path integrals and field equations. For example, a partition function with a source that couples to an operator in the Lagrangian of a quantum field theory is standardly used to calculate, by taking functional derivatives, the correlation functions of that operator in the vacuum state. Cf.~Section \ref{sptthies}, and De Haro et al.~(2017:~pp.~75-76).\label{nontriple}} 

Second:  We admit that of course, a physical theory is almost never presented to us as a tidy triple of state-space, quantities and dynamics. Almost always, the theory appears to us messier than that: more vaguely defined and-or more complicated. The triple needs to be extracted from that ``mess''. Indeed, there are two points here. \\
\indent (i): The complicated appearance is of course in part due to those aspects such as experiment mentioned above. But this complexity, and the need to allow for such aspects (and to assess them philosophically), does not militate against extracting a triple as a concept useful  for e.g.~understanding symmetries.  \\
\indent (ii): We make no claim that there is always, or even typically, a unique best definition of the triple.  So presenting a theory as a triple usually involves: (a) choices about exactly what to take as the state-space etc.; and even (b) judgment about interpretative and perhaps controversial matters. We will see some examples of this variety, already in Section \ref{rootspec};  and we will see that these choices may affect verdicts of interest, e.g.~whether there is an isomorphism, or a duality.

\subsubsection{Interpretation}\label{interp}

This brings us to the interpretation of physical theories. A large subject! But here, we only need to report another part of our overall position, which is again mainstream. Namely, we endorse the endeavour of giving theories (and so: models in our sense) a classical referential semantics. That is: we envisage assigning references in the empirical world to  appropriate  elements of theories---be they states and quantities (i.e.~mathematical objects) for a theory presented as a triple, or words and sentences (i.e.~linguistic objects) for a theory presented in a language. This even-handedness between mathematical and linguistic objects---and thus between theories as $N$-tuples and as sets of sentences---is deliberate: for we endorse recent arguments against the traditional firm distinction between the semantic and syntactic conceptions of theory (Lutz (2017), Glymour (2013), van Fraassen (2014)).\footnote{And our phrase `words and sentences' is to signal that the semantics is compositional in the usual sense: viz.~the reference assigned to a string of symbols is a function of the references assigned to the symbols.}   

This even-handedness is one reason we call our endorsement of referential semantics `mainstream'. There are also two other reasons: the first is familiar in philosophy of science, and the second is familiar in philosophy of language. 

First: recall that accepting such referential semantics is independent, or at least largely independent, of the debate over scientific realism. For that debate is largely about what is the right attitude to our theories, not about their semantic content. Thus the arch empiricist, van Fraassen, explicitly accepts a literal construal of the theoretical claims of---i.e.~a referential semantics for---scientific theories (1980:~p.~14): as do other influential positions that reject realism, such as Fine's `natural ontological attitude' (1984:~pp.~96-99, 1986:~p.~130). 

Second: we should recall the moral of Lewis' seminal paper, `Languages and Language' (1975). Lewis begins by rehearsing a thesis and an antithesis: the task of the paper, and his moral, is to reconcile them in a synthesis---which indeed he accomplishes. Thus the thesis begins by saying that {\em a language} is an assignment `of meanings to certain sequences of types of sound or of marks ...' (p. 3). As this quote hints: the thesis is advocacy of a referential semantics; (indeed an intensional semantics---cf.~below).  The antithesis begins by saying that `{\em language} is a social phenomenon wherein people utter strings of vocal sounds ... and wherein people respond by thought or action to the sounds they observe to have been produced' (ibid). As this quote hints: the antithesis is advocacy of an account of language emphasising people's propositional attitudes (intentions, beliefs, desires etc.): both as what is communicated by language, and as what underpins that communication. Thus Lewis' synthesis is his account of what it is for {\em a} language $L$, {\em \`a  la}  the thesis, to be {\em the} language used by a human population,  {\em \`a  la} the antithesis. The main idea is that this is a matter of the population having conventions (in Lewis' sense) of truthfulness in $L$ and trust in $L$. Thus he knits the thesis and antithesis together in a detailed way (and replies to various objections). He ends by saying: `According to the proposal I have presented, the philosophy of language is a single subject. The thesis and antithesis have been the property of different schools; but in fact they are complementary essential ingredients in any adequate account either of languages or of language' (p.~35).

To which we say: `Hear, hear!'. That is: we claim a similar reconciliation between our advocating a referential semantics for scientific theories, and various lines of philosophical work that downplay, or even do not mention, reference or theories.
Thus some work on some of the aspects of scientific endeavour, mentioned above, seems sceptical, not just about `theory' as a useful notion, but also about reference. For example: much philosophical writing about experiment (calibration of instruments etc.) emphasizes non-linguistic skills, practices and norms; and much philosophical writing about  the social dimensions of knowledge emphasizes the wider and practical world, e.g.~the functioning of scientific communities and institutions in e.g.~the maintenance of norms, such as accreditation, etc. Such emphases are entirely appropriate, say we. Study of experiment should {\em of course} emphasize non-linguistic skills; and so on. But such emphases in no way militate against developing a referential semantics for scientific language, and so for scientific theories. In  short, we think Lewis' synthesis gives a valuable, because irenic, perspective on  this work's relation to referential semantics.\footnote{More controversially: we think Lewis' synthesis gives a valuable, because deflating, perspective on the burgeoning literature about scientific representation; viz.~along the lines  Callender and Cohen's claim that we should analyse representation in science in terms drawn from philosophy of language and mind (2006, especially Section 3).}

So much by way of defending ourselves as having mainstream views about interpretation. The upshot is that we envisage a referential semantics using interpretation maps, $i$. These are maps on states or quantities ($s \in {\cal S}$ etc.), or on linguistic items, assigning as values (outputs of the map) parts of the empirical world (hunks of reality!). But they are, in general, partial maps; i.e.~for some arguments, the map yields no value (output). For some details and examples, see De Haro (2018a).\footnote{In fact, we endorse a specific programme within the general enterprise of referential semantics. Namely: intensional semantics in the sense developed by Carnap, Montague and Lewis in which:\\
\indent \indent (i)  the notion of `linguistic meaning' is taken to be ambiguous between what Frege called `sense' and what he called `reference', here called `intension' and `extension' respectively;   and \\
\indent \indent (ii) intension is modelled as a function taking linguistic items---and for us, states or quantities---to their reference, relative to a possible world.\\
But in this paper, we will not need the details of this view. A standard exposition is Lewis (1970); besides, pp.~16-17 of his (1975) give a fine sketch, including a defence of semantics adverting to possible worlds. Our own endorsement is in: De Haro and Butterfield (2018:~Section 2.3), Butterfield (2018:~Section 3.1), De Haro and De Regt (2018).}\\

In the next Section we will contrast {\em internal} and {\em external} interpretations, both of which are interpretations in the above sense. Roughly speaking, an internal interpretation only interprets that ``part'' of the model that is ``the homomorphic copy'' of the theory, while an external interpretation can interpret all of the model. 

\subsection{Model roots, specific structure, and their interpretations}\label{rootspec}

As hinted at the end of the previous Section, we need some more jargon and notation 
about the relation between a bare theory and its models that will allow us to distinguish the parts or aspects of models that are ``shadows'' of corresponding parts or aspects of the bare theory, from those that are not. 

The jargon is clearest for the common case, which we will focus on throughout this paper: when the realization of the bare theory proceeds by (mathematical) representation, and the bare theory is a triple comprising a state-space $\cal S$, a set of quantities $\cal Q$ and a dynamics $\cal D$: (cf.~Section \ref{theory} and footnote \ref{nontriple}). Although the bare theory may  be interpreted (like its models, the dual theories, usually are), adding the adjective `bare' signals helpfully that the theory is  ``above'' and can be / is often uninterpreted.\footnote{We take the point, from Read and M\o ller-Nielsen (2018:~Section 5.3), that not every common core need be so rich as to lead to a theory (in the present case, not every common core needs to allow being written as a triple). But we will simply restrict our attention to cases in which the common core {\it is} as rich as a theory, which we submit are the most interesting cases---and indeed they seem to be the most common cases in the literature on dualities. See also De Haro (2018a:~\S2.1.1; footnote 30).} 

In this common case: by the very definition of `representation', the model gives a {\em homomorphic copy} of the bare theory (irrespective of there being a duality).  That is: there are appropriate structure-preserving maps from the states, quantities and dynamics of the bare theory to the model's homomorphic copy. To be precise: there is a pair of structure-preserving maps---from states in the bare theory to states in the model, and from quantities in the bare theory to quantities in the model. And there is a meshing condition on the model's dynamics that makes it implement that of the bare theory. The details are as follows: though we can mostly take them in our stride, and just say `homomorphic copy'.\footnote{So the relation of representation between a bare theory and a model of it will involve not just one map as in, say, group representation theory (the homomorphism from the abstract group to e.g.~a set of matrices), but at least two maps. We will see in Section \ref{dualdef} that these two maps are related to each other, because states and quantities are dual (in mathematicians' sense!) to one another.}

We can write the bare theory as a triple $T = \bra {\cal S}, {\cal Q}, {\cal D}  \ket$; and similarly its model $M = \bra {\cal S}_M, {\cal Q}_M, {\cal D}_M  \ket$, where the subscripts signal that the state-space etc.~are different  from that of the bare theory.  If we think of the dynamics in Schr\"{o}dinger style as a map on the state-space (more details in Section \ref{values}), and write the homomorphism from ${\cal S}$ to ${\cal S}_M$ as $h$, then the meshing condition on the model's dynamics will be the commuting diagram in Figure \ref{meshdynamicsbasic}. \\

\begin{figure}
\begin{center}
\bea
\begin{array}{ccc}{\cal S}&\xrightarrow{\makebox[.6cm]{$\cal{D}$}}&{\cal S}\\
~~\Big\downarrow {\sm{$h$}}&&~~\Big\downarrow {\sm{$h$}}\\
{\cal S}_M&\xrightarrow{\makebox[.6cm]{${\cal {D}}_M$}}&{\cal S}_M\nonumber
\end{array}\nonumber
\eea
\caption{The model's (Schr\"{o}dinger) dynamics  implements that of the bare theory.}
\label{meshdynamicsbasic}
\end{center}
\end{figure}

Once we have the distinction of levels, i.e.~a bare theory  represented by a model, there is an important distinction to be made  within the model. (This is important  irrespective of there being a duality.) Namely between:\\
\indent \indent  (i): the parts and aspects of the model which together express its realizing the bare theory;\\
\indent \indent  (ii): the parts and aspects  which do not express the realization. \\
We will call (i) the {\bf model root}. And in the common case where the bare theory, and so also the model, is a triple of states, quantities and dynamics, we will call (i) the {\bf model triple}. And we will call (ii) the model's  {\bf specific structure}.  We can think of it as the `ingredients' or `building blocks' by which the representation of the bare theory, i.e.~(i), is built. 

There is a correlative distinction between two kinds of interpretation. Recall from Section \ref{interp} that an interpretation is given by interpretation maps, i.e.~functions (in general, partial functions) mapping items in the theory  to items in the world. Thus we say that an {\bf internal} interpretation is one that only interprets the model root: it maps all of and only the model root to items in the world, regardless of the specific structure. On the other hand, an {\bf external} interpretation also maps (some or all of) the specific structure to items in the world.

Of course, ingredients are present in the cooked dish, and building blocks are present in the built house. Similarly here: often, an item of specific structure is in the model root, though (by definition) it is not part of the representation of the bare theory. And in such a case an internal interpretation does not interpret the item of specific structure. (Section \ref{explematriGalil} will give examples.)

Thus the distinction between specific structure (`building blocks'), and model root (`what gets built') is physically significant, in that it constrains interpretation. It is formal in that it can be stated without giving an interpretation: but it has consequences for  interpretation. For more discussion, cf.~De Haro (2016:~Section 1.1.2, 2018:~Section 2.2.3), De Haro and Butterfield (2018: Section 3.2.2), and De Haro (2018a:~Section 2.1.2). \\

But we stress that even within a given model, the distinction is not `God-given'. It is relative to how exactly we define the bare theory, and thereby the homomorphism from it to the model. And for a physical theory, as usually presented to us informally and even vaguely, there need be no best or most natural way to make this exact definition. For recall comment (ii) at the end of Section \ref{theory}: how exactly to present a theory as a triple of states, quantities and dynamics is a matter of choice and even judgment. 

We will see this flexibility in play later, in Section \ref{explematriGalil}, where we will have a choice to define the model root either as a single representation of a structure (with a further choice to include or not to include a choice of basis in the model root) or as an equivalence class of representations.

We will also see it in connection with dualities, in Section \ref{expleGalil}. For as we announced: we say a duality is an isomorphism of models of a bare theory; (details in Section \ref{dualdef}). But this means: an isomorphism with respect to the structure of the bare theory---which is the structure that the models represent. Therefore the isomorphism that is the duality is an isomorphism of model roots. And in the common case of triples  of states, quantities and dynamics: it is an isomorphism of model triples.\footnote{A clarification: This is not to say that the specific structure (the `building blocks') is always `invisible' to the other side of the duality, i.e.~that no part of the specific structure is mapped across by a duality to the other model. Often, some part of the specific structure is mapped across. Indeed, that is unsurprising. For the model root  is built from specific structure: so one expects that in order to map the model roots, one into the other, the duality must map at least part of the specific structure, one into the other.\label{notinvisible}} So the flexibility about the definition of a bare theory, and so about what a model must represent, leads to flexibility about exactly what the duality mapping is, i.e.~what is the isomorphism between model roots. \\

It is worth having a notation distinguishing  between the model root (which is usually a model triple of states etc.) and the specific structure. So given a model $M$, we now write $m$ for its model root. This is usually a model triple, which we now write as $\bra{\cal S}_M,{\cal Q}_M,{\cal D}_M \ket$. Again, the subscripts signal that the state-space etc.~are different from those of the bare theory. We also write $\bar M$ for the specific structure.  So we write a model $M$ of a bare theory $T$ as
\beq\label{eqmodel}
 M  =:\bra m ; \bar M\ket = \bra{\cal S}_M,{\cal Q}_M,{\cal D}_M ; \bar M\ket~,
\eeq
 where the semi-colon in the defined angle-bracket signals the contrast between root and  specific structure, and the last equation just expresses the usual case of the root being a triple.

But {\em beware}: one should not think of $M$ as just the ordered pair of independently given $m$ and $\bar M$. For $m$ is  built by using the specific structure $\bar M$, and so it is not given independently of $\bar M$. Rather, $m$ encodes $M$'s representing the bare theory $T$. So one might well write $T_M$ instead of $m$, since having a subscript $M$ on the right hand side of the equation $M = \bra T_M; \bar M \ket$ signals that $M$ is not  an ordered pair of two independently given items. In other words: the notation $T_M$ emphasises that the model triple: (i) is a representation of $T$, (ii) is built from $M$'s specific structure viz.~$\bar M$, and (iii) is itself a theory (hence the letter `T').

Here is an illustration of this notation in use. As announced: we say that a duality is an isomorphism of model roots, with respect to the structure of the bare theory. So if $M_1, M_2$ are models of a bare theory $T$, their being duals means: $m_1 \cong m_2$. And in the usual case of model triples, i.e.~$m_i =  \bra{\cal S}_{M{_i}},{\cal Q}_{M{_i}},{\cal D}_{M{_i}} \ket\,,~i = 1,2$: this isomorphism of roots will be a matter of two appropriately meshing isomorphisms, one between the state-spaces ${\cal S}_{M{_i}}$ and  one between the quantity algebras ${\cal Q}_{M{_i}}$. Details in Section \ref{dualdef}. 
 
\subsection{Examples: matrix representations and Galilean  transformations}\label{explematriGalil}

In this Section, we will illustrate our notions of theory and model, and the contrast of model root vs.~specific structure. Our first example (Section \ref{explematri}) is from matrix representations; our second example (Section \ref{expleGalil}) is about Galilean transformations in Newtonian mechanics. The latter example will also illustrate our notion of interpretation from Section \ref{interp}, especially our internal vs.~external contrast from Section \ref{rootspec}. 

\subsubsection{Matrix representations}\label{explematri}

 Perhaps the simplest illustration of these notions, model root and specific structure, comes from defining the bare theory to be just an abstract group $G$; and as usual, taking realizations to be representations.\footnote{Strictly speaking, a bare theory must have an appropriate set of maps to the real numbers, to express the values of quantities, and even a dynamics. But for the sake of illustration, we here ignore these maps. Anyway, Section \ref{expleGalil} will sketch an illustration from physics.} So let us consider matrix representations of $G$. More specifically: we consider for a finite group $G$, a set $\{ M_i \}$ of $n \times n$ complex matrices with non-zero determinant ($i$ runs from 1 to the order of $G$). That is: $ M_i \in \mbox{GL}(n,\mathbb{C})$. In such a representation,  the choices of the size $n$ of the matrices and of the ground-field ($\mathbb{R}$ or $\mathbb{C}$) are of course parts of specific structure: for these are building blocks by which we build the homomorphic copy of $G$. 
 
But even in this simple illustration, we can see further options about how exactly to define model root and specific structure. These options relate to the fact that the unitary group $\mbox{U}(n)$ acts  on $\mbox{GL}(n,\mathbb{C})$, by $M \mapsto U M U^{-1} \equiv U M U^{\dagger}$. Each $U \in \mbox{U}(n)$ sends a representation $\{ M_i \}$ to another representation, called `equivalent'; and representation theory then of course concentrates on equivalence classes of representations, characterizing them in terms of their invariants, especially characters.  The spirit of the enterprise is that a unitary change of basis has no mathematical significance. This means there are two main options, (A) and (B) below, about how to define the model root, and thereby, also the specific structure---and there will be further choices.\footnote{Roughly speaking, (B) takes roots to be equivalence classes of what (A) takes them to be. One might question whether (A)'s ``concrete'' matrices are really more ``basic'' than their equivalence classes as in (B). But exposition of the issues is much clearer if we keep the (A) vs.~(B) contrast. }\\
 
 (A): {\em Model root as representation}:--- We can say that a single representation $\{ M_i \}$ is the model root. Indeed: multiplication of the matrices $M_i$ realizes a homomorphic copy of $G$'s multiplication. And it is no objection that items of specific structure---the size of the matrices $n$, and the choice of the ground-field---are `in' the matrices $M_i$. For as we said: that an item appears in a model root does not prevent it appearing in the specific structure. After all, the model root is {\it built from} the specific structure.
 
 On this option, two equivalent representations of $G$, i.e.~a set of matrices $\{ M_i \}$ and another set $\{ U\,M_i\,U^{-1}  \}$, for some fixed $U \in \mbox{U}(n)$, will be isomorphic, as homomorphic copies of $G$'s multiplication.  So  on our account of duality, they are duals. (As we will discuss in Section \ref{schema}: this reflects our definition of duality being logically weak i.e.~having many instances.)  
 
 But there is also a further choice. For the fact that the matrices realize group multiplication is independent of their acting as linear operators on the vector space $\mathbb{C}^n$. That is: although we always think of a matrix as representing (that word again!) a linear operator, it only does so relative to a choice of basis---and the representation we began with, viz.~$h: G \rightarrow \{ M_i \}$, makes {\em no} such choice. But if we wish, we can adjoin such a choice to our model. That is: we  can stipulate that the matrix representation $\{M_i\}$ also has, as part of its specific structure, some specific basis ${\bf e}_k$ ($k=1,\ldots,n$) of $\mathbb{C}^n$.  
 
 Once such a choice is made, the basis vectors ${\bf e}_k$ (and thereby all vectors ${\bf v}=\sum_{k=1}^nv_k\,{\bf e}_k$) are of course acted upon---sent to another basis---by the similarity transformations $U$ that act on the matrices $M_i$. That is: ${\bf e}_k\mapsto \sum_{l=1}^n U_{kl}\,{\bf e}_l$ when $M_i\mapsto U\,M_i\,U^{-1}$. But the idea of the stipulation is that the first-chosen basis ${\bf e}_k$ labels the representation: it fixes the interpretation of the $\{M_i\}$ as linear operators. It is just that this labelling basis maps across to the labelling bases of equivalent representations. This illustrates the idea in footnote \ref{notinvisible} that---to return to our jargon---specific structure can map across a duality. In this example, the duality maps a labelling basis to another basis: $\{ {\bf e}_k \}\mapsto \{ \sum_{l=1}^n U_{kl}\,{\bf e}_l \} $ when $M_i\mapsto U\,M_i\,U^{-1}$. \\

  (B): {\em Model root as equivalence class}:--- We can say that an equivalence class of representations---the entire orbit of a given $\{ M_i \}$ under the action of $\mbox{U}(n)$---is the model root. For indeed: equivalent representations realize the very same homomorphic copy of $G$'s multiplication. Again, the size of the matrices $n$, and the choice of the ground-field, are `in' the model root (since they are preserved by equivalence). And again, it is no objection that an item in the model root is also in the specific structure: since the model root is built from the specific structure.
  
On option (B), there are still duals,  on our account of duality---despite the option's having ``quotiented'' to a more abstract notion of model than that of option (A). For we can vary $n$; and-or we can vary the ground-field.  That is: two different choices of $n$ can provide two model roots---the equivalence class of a representation $\{M_i\}: M_i \in \mbox{GL}(n_1,\mathbb{C})$, and the equivalence class of a representation $\{M_j\}: M_j \in \mbox{GL}(n_2,\mathbb{C})$---that instantiate the same homomorphic copy of $G$. And similarly, we can vary the ground-field, and yet instantiate the same homomorphic copy of $G$. And similarly, we can  consider non-linear/non-matrix realizations/representations of $G$.

And as in option (A), there is the further choice---again because the fact that matrices realize group multiplication is independent of their acting as linear operators on the vector space $\mathbb{C}^n$. That is: if we wish, we can adjoin to a model root---an entire equivalence class of representations---the choice of some specific basis ${\bf e}_k$ ($k=1,\ldots,n$) of $\mathbb{C}^n$, which we can then take to be specific structure. Of course, the only natural way to do this is to attach the basis to some arbitrary element of the class, i.e.~one set of matrices $\{ M_i \}$, and then transport the basis around to the other elements of the equivalence class by the action of $\mbox{U}(n)$.

\subsubsection{Galilean  transformations}\label{expleGalil}

Newtonian mechanics provides a simple illustration of the notions of model root, specific structure, and indeed duality. And since it is an example from physics rather than pure mathematics, we also get an illustration of Section \ref{rootspec}'s distinction between internal and external interpretations.

The idea is as follows. The bare theory $T$ is Newtonian mechanics, of say $N$ gravitating point-particles, set in a Galilean (neo-Newtonian) spacetime: i.e.~in a spacetime manifold that is `globally like $\mathbb{R}^4$', with Euclidean geometry in its instantaneous time-slices, and a flat 4-dimensional connection, but no preferred absolute rest. This bare theory is modelled (in our sense: i.e.~realized, represented) by formulations of Newtonian mechanics of $N$ gravitating particles, set in a {\em Newtonian} spacetime, i.e.~in a spacetime that is `globally like $\mathbb{R}^4$' but that {\em does} have an absolute rest. 

Famously (notoriously!), the difference in such formulations' specifications of absolute rest is not experimentally detectable, since specifications that are each boosted with respect to the other specify the same flat 4-dimensional connection, and a boost maps a solution of the equations of motion to another solution. Or as it is usually put, in more physical terms: no experiment in Newtonian mechanics can distinguish one specification of rest from the others, because the theory is invariant under boosts (`has boosts as a symmetry'). Hence, of course, the debate between Newton and Leibniz, as articulated in the Leibniz-Clarke correspondence, and with its long legacy down to the present day (e.g.~Earman 1989).

 So in this example, it is natural to say that the specific structure of each model includes its specification of absolute rest. Using this  specification, the model defines a flat 4-dimensional connection---viz.~the same connection as is defined by the other models---and thereby builds a homomorphic copy of $T$. 
 
We can make this example simpler and precise, and yet still a worthwhile illustration, by taking the bare theory $T$ to be just the abstract Galilean group Gal(3). This is a 10-dimensional Lie group, whose generators are three spatial rotations, four (space and time) translations, and three boosts. That is: its generators are usually thus described, by way of justifying their commutation relations. But  of course the abstract group can be defined by the commutation relations, free of a physical interpretation. Thinking of Gal(3) like this will yield a clear analogy with Section \ref{explematri}'s matrix representations of an abstract group $G$, and with that Section's option (A), i.e.~{\em Model root as representation}.\\
 
Gal(3) is usually presented in its fundamental representation. Namely, as a concrete group of  transformations on (bijections of) $\mathbb{R}^4$, written in terms of the standard coordinates $({\bf x},t) \in \mathbb{R}^4$, with $g \in$ Gal(3) represented as the function
\bea\label{galilean}
g({\bf x},t):=(R\cdot {\bf x}+{\bf v}_0\,t+{\bf r}_0,t+t_0)~,
\eea
where $R$ is a $3\times3$ spatial rotation matrix, ${\bf v}_0$ is the velocity boost, ${\bf r}_0$ is the spatial translation vector, and $t_0$ is the time translation.  This fundamental representation can also be expressed in a coordinate-free way as an action on the affine space of $\mathbb{R}^4$, i.e.~on Euclidean 4-space. But we will not need the details of affine spaces (cf.~e.g.~Auslander and MacKenzie (1963, Chapter 1)).

But just as in Section \ref{explematri}'s option (A), where a model root was a matrix representation of a group $G$, we could adjoin a choice of a basis ${\bf e}_k$ ($k=1,\ldots,n$) of $\mathbb{C}^n$ as specific structure: so also here, we can adjoin a choice of an inertial coordinate system as  specific structure, and we can take this to give the model's specification of absolute rest.\footnote{We say `give'---meaning `determine'---rather than `be', simply because a coordinate system includes choices of spatial and temporal origins and units, and of an orientation of spatial axes, as well as the absolute rest, i.e.~the timelike congruence of inertial worldlines.\label{coordcong}}  The standard coordinate system defined by the components of $\mathbb{R}^4$ itself is then just one choice among many, determining one specification of absolute rest among many. Natural though we find it for writing down the fundamental action of Gal(3), as we did in Eq~\eq{galilean}, the action can of course be written down in any inertial coordinate system. And any such system can be taken to give a model's specification of absolute rest. 

And just as in (A) of Section \ref{explematri}, each matrix $M_i$ mapped the adjoined basis ${\bf e}_k$ to another basis: so also here, each Galilean boost maps an adjoined choice of absolute rest, represented mathematically by an inertial coordinate system, into another such choice, i.e.~another inertial coordinate system.\footnote{Of course, Galilean  transformations that are not boosts keep fixed the choice of absolute rest: cf.~footnote \ref{coordcong}.} 

But there is also, so far, a disanalogy with (A) of Section \ref{explematri}. For so far, we have just one model root, encapsulated in  Eq.~\eq{galilean}, and various choices of specific structure; whereas (A) of Section \ref{explematri} had many different model roots---many different matrix representations $\{ M_i \}$ of the abstract group $G$. Correlatively, our fundamental representation of Gal(3) is faithful, i.e.~has trivial kernel; while Section \ref{explematri}'s $\{ M_i \}$ were in general not faithful.

But there are other representations of Gal(3). Indeed, there is a {\em matrix} representation that is isomorphic, as a model root, to what we have. So it is faithful---and the isomorphism of model roots is, on our account, a duality. Namely, we use $5\times5$ matrices $\mbox{G}\in\mbox{GL}(5,\mathbb{R})$ ($\mbox{G}$ for `Galileo' not `group'!), that act on vectors $(x_A):=({\bf x},t,1)\in\mathbb{R}^4\times\{1\}\subset\mathbb{R}^5$, as follows (cf.~Bargmann (1954:~pp.~38-41) and Holm (2011:~p~10)):
\bea\label{bargm}
\ti x_A&=&\sum_{B=1}^5\mbox{G}_{AB}\,x_B\\
(\mbox{G}_{AB})_{{A,B}=1}^5&:=&\left(\begin{array}{ccc}R&{\bf v_0}&{\bf r}_0\\{\bf 0}_{1\times3}&1&t_0\\{\bf 0}_{1\times 3}&0&1\end{array}\right).\nonumber
\eea
With this representation, a specification of absolute rest is given by a specific choice of vector $x\in\mathbb{R}^4\times\{1\}$. Namely, $(x_A):=({\bf x},t,1)\in\mathbb{R}^4\times\{1\}$ specifies rest to be the timelike congruence of inertial worldlines parallel to the inertial worldline passing through both the spatiotemporal origin $({\bf 0}, 0)$ and the point $({\bf x},t)$. So here, it is a single vector that gets adjoined as specific structure; (not a whole basis, as in Section \ref{explematri}, and not a whole inertial coordinate system, as above). And  each Galilean boost maps an adjoined choice of absolute rest, given by a vector $x\in\mathbb{R}^4\times\{1\}$, into another such choice, i.e.~another vector.

So much by way of how Galilean  transformations' illustration of the notions of model root and specific structure is analogous with Section \ref{explematri}'s matrices. We end with two comments, [1] and [2], about the physical interpretation of this example (comments which thereby have no analogues about those matrices). For simplicity and brevity, we will restrict both comments to the simple ``vacuum'' scenario which we have concentrated on: i.e.~$\mathbb{R}^4$ as a description of either neo-Newtonian or Newtonian spacetime, without regard to the $N$ gravitating point-particles we mentioned at the start of this Section. Thus recall that we concentrated on taking  the bare theory $T$ to be just the abstract Galilean group Gal(3), and considered its action on $\mathbb{R}^4$. But this is only for simplicity: these two comments carry over to the non-vacuum  scenario, where there are particles.\footnote{It is just that it would take too long to spell out the non-vacuum  scenario. To glimpse why, we briefly note some of the issues one confronts. Obviously, one  must consider the particles' state-space: which one would build from their configuration space (the ``$q$s''), by adding either velocities (``$\dot q$s'': defining velocity phase space, in the Lagrangian framework)  or canonical momenta (``$p$s'': defining phase space, in the Hamiltonian framework). At first sight, the $N$ particles' configuration space is ``just'' $\mathbb{R}^{3N}$. But there are subtleties to be dealt with. Indeed: not only the topics mentioned above, of passing to the affine space so as to ``rub out'' the preferred origin, and whether to have an absolute rest; but also whether to excise collision points, i.e.~whether to forbid point-particles to be in the very same place.  Assuming these subtleties are dealt with, and the Lagrangian or Hamiltonian state-space is constructed, one would then consider the action of the Euclidean group on this state-space, lifted from its action on  $\mathbb{R}^3$. Again, there are subtleties about this lifted action; and to treat boosts and so represent the Galilean group, one needs to ``add a time axis'', defining what is often called `extended (velocity) phase space'. For a philosophical introduction to all these subtleties, cf.~e.g.~Belot (2000: Sections 3,4) and Butterfield (2006: Section 2.3).

These two comments also bear on the two debates about interpreting symmetries, which we set aside at the end of Section \ref{intro}. Indeed the authors cited take Newtonian  gravitation as a main example. But again, these debates are  orthogonal to most of this paper's issues; so that it would take too long to spell out, beyond the references we gave there, exactly what these comments imply for them.\label{subtleties}}  \\

[1]:  {\em Agreement with usual verdicts}:--- The first comment looks ahead to Section \ref{sptthies}'s discussion of symmetries in a spacetime theory, i.e.~a theory that postulates a spacetime with certain chrono-geometric structures like metrics and connection. (Such theories of course also postulate matter and radiation, particles and fields, in the spacetime; but as just announced, we are setting that aside.) 

We will see in Section \ref{sptthies} that in a spacetime theory, a symmetry is usually defined as a map on the spacetime that (once its domain of definition is extended in the natural way to include chrono-geometric structures and matter fields): (i) fixes, i.e.~does not alter, the chrono-geometric structures, and (ii) maps a matter-field solution of the equations of motion to another solution. (We will also see how this relates to the more basic and general notion of symmetry we will use from the start of Section \ref{symmygenl}.) Accordingly, boosts are a symmetry of neo-Newtonian spacetime: for a boost preserves the chrono-geometric structures, i.e.~the spatial and temporal metrics and the flat affine connection, and maps solutions to solutions. But boosts are not a symmetry of Newtonian spacetime (i.e.~a spacetime that is globably like $\mathbb{R}^4$, but that has a specification of absolute rest). For a boost does not fix a specification of rest.\footnote{On the other hand: spatial rotations and spatiotemporal translations are symmetries of both spacetimes.} These points are, in effect, the modern mathematical expression of the famous (notorious!) point with which this Section began: that no mechanical experiment can discern which is the putatively correct standard of rest. 

 Our discussion above, and our notions of model root, duality etc., accords with this. In particular, just as in Section \ref{explematri}: an item of specific structure can be ``in'' the model root, which is, after all, built with specific structure; and so an isomorphism of model roots (on our account, a duality) can map specific structure from one model to another. And this is what Galilean boosts do. In our jargon: they define an isomorphism of model roots that maps one model's specification of absolute rest into another's. Think for example of how the vector  $(x_A):=({\bf x},t,1)\in\mathbb{R}^4\times\{1\}$---which specifies rest as parallelism to the inertial worldline through both the origin $({\bf 0}, 0)$ and the point $({\bf x},t)$---is mapped by $\mbox{G}_{AB}$ of Eq.~\eq{bargm} to a vector specifying rest as parallelism to the inertial worldline through both the origin $({\bf 0}, 0)$ and the point $(R\cdot {\bf x}+{\bf v}_0\,t+{\bf r}_0, t+t_0)$.
(For more detail about how the duality maps in this example are defined by boosts, cf.~Butterfield (2018:  Section 4.1).)

But mapping one model's specification of absolute rest into another's is not the same as {\em fixing} the given specification, i.e.~not the same as a symmetry in spacetime theories' usual sense. Thus this example illustrates how a bare theory can have a symmetry, viz.~boosts, that (some or even all) its models lack. This will later be a main theme (Sections  \ref{secure} and \ref{fail}).

 This example also has a philosophical moral that is not about symmetry. Namely: duality does {\em not} imply physical equivalence. Two theories can be duals---in our jargon: models with isomorphic model roots---without their making the very same claims about the world. They can even contradict one another about the world, as do two rival specifications of what is absolute rest. This leads in to the next comment.\\

[2]:  {\em Internal and external interpretations}:---  The example also illustrates Section \ref{rootspec}'s distinction between internal and external interpretations. As usual, interpretative issues are underdetermined by formal theory: witness the moral just stated at the end of [1].   However, the example clearly allows us to formulate the disagreement between Newton (Clarke) and Leibniz---the question whether absolute rest is physically real---in terms of the internal vs.~external contrast. 

For recall that an internal interpretation interprets only the model root, but not the specific structure. More precisely, we define this as meaning that specific structure which is ``in'' a model root as a building block, does not get interpreted. Therefore models that are isomorphic, i.e.~have isomorphic roots,  and so differ only in their specific structure, must receive the same interpretation.\footnote{Recall that Newtonian gravitation contains unphysical singularities when two massive point particles coincide, as they can do after a finite time (see footnote \ref{subtleties}). Thus internal interpretations of isomorphic model roots of Newtonian gravitation must either give the same ``interpretation'' to the unphysical singularities (usually: signaling a limitation of the applicability of the theory), or one must deal with the singularities in some other way. See the discussion in De Haro (2018a:~Sections 2.3.1, 2.3.2(C)).} Thus in our example: an internal interpretation of a model simply does not interpret the specification of rest (whether it is given by an inertial coordinate system considered as rest, as for equation \eq{galilean}, or by a vector $x\in\mathbb{R}^4\times\{1\}$, as for equation \eq{bargm}). In short: the specification of rest  is not part of what is physical. Thus an internal interpretation articulates Leibniz' relationist views.

 On the other hand, an external interpretation  (by definition) does interpret (some or all of) the specific structure. Thus an external interpretation can take any two dual models---any isomorphic model roots with different specifications of rest, $x$ and $\ti x$ say---to have distinct interpretations. This kind of external interpretation thus articulates the Newton-Clarke view: in short, that giving all material bodies the same boost makes a physical difference.

 \subsection{Values of quantities on states}\label{values}
Before formally defining duality as isomorphism, we need notation for treating states, quantities and dynamics. Suppose we are given a set of states ${\cal S}$, a set of quantities ${\cal Q}$ and a dynamics ${\cal D}$: $\bra {\cal {S}}, {\cal {Q}}, {\cal {D}} \ket$. (As stressed in Section \ref{intro}: in any specific case, ${\cal S}$,  ${\cal Q}$ and ${\cal D}$ will each have a lot of structure beyond being sets---but we will not need these details  in what follows. And as admitted in Section \ref{theory}: not all theories are presented, or best thought of, as such triples---but what we say will carry over to other formulations using e.g.~partition functions; cf.~footnote \ref{nontriple}.) 

Then we write the value of quantity $Q$ in state $s$ as
\be\label{pairing}
\langle Q , s \rangle
\ee
It is these values that are to be  preserved, or suitably matched, by the duality, i.e.~by the isomorphism of model triples: cf.~Eq.~\eq{obv1} below. And in subsequent Sections' discussion of symmetry, it is these values that are to be preserved by a symmetry map.
 
For classical physics, one naturally represents (that word again!) a quantity as a real-valued function on states: $Q: s \mapsto Q(s)$. Given such a function representing the quantity, $\langle Q , s \rangle := Q(s) \in \mathbb{R}$ 
 is the system's possessed or intrinsic value, in state $s$, of the quantity $Q$. Similarly for quantum physics: one naturally represents quantities as linear operators on a Hilbert space of states, so that  $\langle Q , s \rangle := \langle s |{\hat Q} | s \rangle \in \mathbb{R}$ is the system's Born-rule expectation value of the quantity. (But in fact, for quantum dualities, the duality often preserves off-diagonal matrix elements $\langle s_1 |{\hat Q} | s_2 \rangle \in \mathbb{C}$: cf.~below.)  
 
 As to dynamics, here assumed deterministic:--- This can be usually presented in two ways, for which we adopt the quantum terminology, viz.~the `Schr\"{o}dinger' and `Heisenberg' pictures; (though the ideas occur equally in classical physics). We will adopt for simplicity the Schr\"{o}dinger picture. So we say: $d_{\cal S}$ is an action of the real line $\mathbb{R}$ representing time on ${\cal S}$. 
 
 There is an equivalent Heisenberg picture of dynamics with $D_H$,  an action of  $\mathbb{R}$ representing time on ${\cal Q}$. The pictures are related by, in an obvious notation:
\be\label{dyns1}
d_{\cal S}:   \mathbb{R} \times {{\cal S}} \ni (t,s) \mapsto d_{\cal S}(t,s) =: s(t) \in {{\cal S}} \;  {\mbox{iff}} \; 
D_H:   \mathbb{R} \times {{\cal Q}} \ni (t,Q) \mapsto D_H(t,Q) =: Q(t) \in {{\cal Q}} 
\ee 
where for all $s \in {{\cal S}}$ considered as the initial state, and all quantities $Q \in {{\cal Q}}$, the values of physical quantities at the later time $t$ agree in the two pictures:
\be\label{dyns2}
\langle Q, s(t) \rangle = \langle Q(t), s \rangle \; .
\ee 

\section{The Schema: Duality as Isomorphism of Model Roots}\label{schema} 

In this Section, we summarize our account of duality.  This  account has been developed mainly by  De Haro (2016:~Section 1, 2016a:~Section 1), but also more fully by us together (2018: Sections 2, 3).
As a mnemonic, we label this account a {\em Schema}. We first define duality (Section \ref{dualdef}); then we give a prospectus for the following Sections (Section \ref{prospectus}). It will be clear that our Schema is logically weak, so that there are countless examples, including elementary ones: a topic taken up in footnote \ref{logiweak}'s references.

\subsection{Duality defined}\label{dualdef}

We can now present our Schema for duality as an isomorphism between model roots (model triples). Let $M_1, M_2$ be two models, with model  roots $m_1$ and $m_2$ and specific structure $\bar M_1$ and $\bar M_2$; so that, with the notation Eq.~\eq{eqmodel}, we have: $M_1=\bra m_1; \bar M_1\ket$ and $M_2=\bra m_2;\bar M_2\ket$. We can suppose that  $M_1, M_2$ are both models of a bare theory $T$. Then we say that $M_1$ and $M_2$ are dual if their model roots are isomorphic, i.e.~if $m_1\cong m_2$.

More specifically, if the model roots are  triples $m_1 =  \bra {\cal S}_{M_1}, {\cal Q}_{M_1}, {\cal D}_{M_1}\ket$ and $m_2 =  \bra {\cal S}_{M_2}, {\cal Q}_{M_2}, {\cal D}_{M_2}\ket$, then to say that the model triples $m_1, m_2$ are isomorphic is to say, in short, that: there are isomorphisms between their respective state-spaces and sets of quantities, that \\
\indent \indent (i) make values match, and \\
\indent \indent (ii) are equivariant for the two triples' dynamics (in the Schr\"{o}dinger and Heisenberg pictures, respectively).\footnote{Alternatively, in other formalisms (cf.~footnote \ref{nontriple}), the dynamics in one model is mapped into the dynamics in the other. For an example, where the dynamics is a set of covariant Euler-Lagrange equations, see De Haro (2018a:~Section 3).}\\
Thus these maps are our rendering of the correspondence between duals: of, in physics jargon, `the dictionary'.\\

Thus we say: A {\bf duality} between $m_1 =  \bra {\cal S}_{M_1}, {\cal Q}_{M_1}, {\cal D}_{M_1}\ket$ and $m_2 =  \bra {\cal S}_{M_2}, {\cal Q}_{M_2}, {\cal D}_{M_2}\ket$ requires: \\
\indent \indent (1): an isomorphism between the state-spaces (almost always: Hilbert spaces, or for classical theories, manifolds): 
\be\label{obv-1}
d_{\cal S}: {{\cal S}_{M_1}} \rightarrow {{\cal S}_{M_2}} \;\;  {\mbox{using $d$ for  `duality'}} \; ; \; {\mbox{and}}
\ee 
\indent \indent (2): an isomorphism between the sets  (almost always: algebras) of quantities\\
\be\label{obv0}
d_{\cal Q}: {{\cal Q}_{M_1}} \rightarrow {{\cal Q}_{M_2}} \;\;  {\mbox{using $d$ for `duality'}} \; ;
\ee 
such that: (i) the values of quantities match: 
\be\label{obv1}
\langle Q_1, s_1 \rangle_1 = \langle d_{\cal Q}(Q_1), d_{\cal S}(s_1) \rangle_2 \; , \;\;~ \forall Q_1 \in {{\cal Q}_{M_1}}\,,~s_1 \in {{\cal S}_{M_1}}. 
\ee
and: (ii) $d_{\cal S}$ is equivariant for the two triples' dynamics, $D_{S:1}, D_{S:2}$, in the Schr\"{o}dinger  picture; and 
$d_{\cal Q}$ is equivariant for the two triples' dynamics, $D_{H:1}, D_{H:2}$, in the  Heisenberg picture: see Figure \ref{obv2}.

\begin{figure}[ht]
\begin{center}
\bea
\begin{array}{ccc}{\cal S}_{M_1}&\xrightarrow{\makebox[.6cm]{$\sm{$d_{\cal S}$}$}}&{\cal S}_{M_2}\\
~~\Big\downarrow {\sm{$D_{S:1}$}}&&~~\Big\downarrow {\sm{$D_{S:2}$}}\\
{\cal S}_{M_1}&\xrightarrow{\makebox[.6cm]{\sm{$d_{\cal S}$}}}&{\cal S}_{M_2}
\end{array}~~~~~~~~~~~~
\begin{array}{ccc}{\cal Q}_{M_1}&\xrightarrow{\makebox[.6cm]{$\sm{$d_{\cal Q}$}$}}&{\cal Q}_{M_2}\\
~~\Big\downarrow {\sm{$D_{H:1}$}}&&~~\Big\downarrow {\sm{$D_{H:2}$}}\\
{\cal Q}_{M_1}&\xrightarrow{\makebox[.6cm]{\sm{$d_{\cal Q}$}}}&{\cal Q}_{M_2}
\end{array}\nonumber
\eea
\caption{Equivariance of duality and dynamics, for states and quantities.}
\label{obv2}
\end{center}
\end{figure}

Eq.~\eq{obv1} appears to favour $m_1$ over $m_2$; but in fact does not, thanks to the maps $d$ being bijections.\\

This definition of duality can be simplified, since the requirement that the values of quantities match, Eq.~\eq{obv1}, implies relations between the duality maps $d_{\cal S}$ and $d_{\cal Q}$.  Thus in the quantum case, the duality maps are related by: $d_{\cal Q}(Q)=d_{\cal S}\circ Q\circ d_{\cal S}^{-1}$ (where $d_{\cal S}$ is constrained to be unitary).\footnote{The proof is as follows. As mentioned in Section \ref{values}, Eq.~\eq{obv1} generalizes to matrix elements between arbitrary vector states $s, s'$ in the state-space, which in the quantum case is a Hilbert space. Namely, the duality maps must satisfy: $\bra s'|Q|s\ket_1=\bra d_{\cal S}(s')|d_{\cal Q}(Q)|d_{\cal S}(s)\ket_2$. Notice that we can rewrite this purely as an identity in model $m_1$:
$\bra s'|\,Q\,|s\ket_1=\bra s'|\,d_{\cal S}^\dagger\circ d_{\cal Q}(Q)\circ d_{\cal S}\,|s\ket_1$. 
Since the state-space is the whole Hilbert space, ${\cal S}_1={\cal H}_1$, this equation is valid for all $s,s'\in{\cal H}_1$ (it is also valid for all $Q\in{\cal Q}$); so it implies that the {\it operators} on the two sides  must be equal:
$d_{\cal S}^\dagger\circ d_{\cal Q}(Q)\circ d_{\cal S}=Q~\Rightarrow~d_{\cal Q}(Q)=d_{\cal S}\circ Q \circ d_{\cal S}^\dagger$, 
as claimed in the main text.  
\label{dQdS}} Thus duality comes down to a {\it single} duality map on states, $d_{\cal S}$, together with appropriate equivariance conditions on the quantities and the dynamics.  

Similarly in the classical  case: though representing quantities as real-valued functions on the state-space, rather than as maps on the state-space, means that the relation between the duality maps $d_{\cal S}$ and $d_{\cal Q}$ is a little different. Here, we need the notion of a dual map (`dual' in the mathematical, not physical, sense). Thus recall that given any map between two sets $f: X \rightarrow Y$, and any map $g: Y \rightarrow Z$, the {\em dual} (or {\em transpose}) $g^*$ of $g$ (with respect to $f$) is defined as the map $g^*: X \rightarrow Z$ with $g^*(x) := (g \circ f) (x)$. Putting $X = Y = {\cal S}$,  $f = d_{\cal S}$, and $Z = \mathbb{R}$, and taking the function $g:  Y \rightarrow Z$ to be the quantity (as a real-valued function) $Q: {\cal S} \rightarrow   \mathbb{R}$: this definition of the dual map becomes (with the bra-ket notation now meaning values of quantities as in Eq.~\eq{pairing}): $\bra Q^*, s \ket := \bra Q , d_{\cal S}(s) \ket$. But so far, the notation $Q^*$ does not exhibit its dependence on $d_{\cal S}$; (just as $g^*$ does not exhibit its dependence on $f$). So it is clearer to write $d_{\cal S}^*(Q)$ instead of just $Q^*$. Thus we write: $\bra d_{\cal S}^*(Q), s \ket := \bra Q , d_{\cal S}(s) \ket$. Applying this to $d^{-1}_{\cal S}: {\cal S} \rightarrow {\cal S}$, we deduce that defining $d_{\cal Q}$ as the dual $(d^{-1}_{\cal S})^*$ of $d^{-1}_{\cal S}$ yields Eq.~\eq{obv1}, as desired. That is: we have by the definition of `dual map' :
\be
\bra (d^{-1}_{\cal S})^*(Q) \, , \, d_{\cal S}(s) \ket =  \bra (Q \circ d^{-1}_{\cal S}), d_{\cal S}(s) \ket = \bra Q, s \ket \; ;
\ee
which is Eq.~\eq{obv1}. So like in the quantum case: duality comes down to a {\it single} duality map on states, $d_{\cal S}$, with $d_{\cal Q}$ being defined as the dual, i.e.~transpose, of its inverse $d^{-1}_{\cal S}$, and appropriate equivariance conditions on the  dynamics.

We will see in Section \ref{symmygenl} that  for symmetries instead of dualities, we can similarly concentrate on a map on states---unsurprisingly, since the basic notion of symmetry is the preservation of the values of given quantities. 

\subsection{The road ahead: duality as a `giant symmetry'}\label{prospectus}

As discussed in Section \ref{intro}, we have elsewhere related this Schema to various topics, and illustrated it with bosonization and some   examples from string theory. The job of the next three Sections is to relate it to  symmetries. 

As also discussed in Section \ref{intro}, it is agreed in the literature that a duality is like a `giant symmetry': a symmetry between theories. The main new ingredient that the Schema adds to this agreed idea is its picture of two levels, with the bare theory above and the model roots, the bare theory's homomorphic copies, below. As we will see, these two levels can differ in the amount of structure that a map, such as a symmetry, is required to preserve; and this prompts some distinctions between types of symmetry. 

Thus we will proceed in three stages. \\
\indent (i): We begin with comments about symmetry in general (Section \ref{symmygenl}). They are regardless of both: (a) there being a duality; and (b) the distinction between a theory and its representations (homomorphic copies), or more generally its instantiations. These comments are familiar ground in the philosophy of symmetry: but they are worth making since they will apply,  suitably adjusted, to the rest of the paper. \\
\indent (ii): Then we discuss, regardless of there being a duality, how a symmetry of a theory is related to symmetries of its representations (homomorphic copies), or more generally its instantiations:  (Section \ref{symmths}). This will yield the distinctions between types of symmetry. \\
\indent  (iii): Finally, we suppose we have  a duality in the sense of the Schema, and relate this to symmetries. That is: we show that a duality preserves the symmetries of its model triples (Section \ref{reltosymmy}).

\section{Symmetries in General}\label{symmygenl}

We begin with a usual notion of symmetry: as a map $a$ on states, $a: {\cal S} \rightarrow {\cal S}$, that preserves the values of a salient set of quantities: usually a large set, though not necessarily all the quantities. The map $a$ must also respect the structure of ${\cal S}$, e.g.~topological or differential structure. (Thus `$a$' is for `automorphism'.) But this requirement will be in the background in the sequel: the emphasis will be on the state $s$ and the image-state $a(s)$ having the same values for quantities in the salient set. 

Here, the notion of value is exactly as in the Schema:  $\langle Q , s \rangle$, understood as a classical possessed value or a quantum expectation value (or more generally, as a matrix element, $\langle s' |\,{\hat Q}\, | s \rangle \in \mathbb{C}$: see below). The equality of values, for a symmetry $a$,
\be\label{symbasic}
\langle Q , a(s) \rangle = \langle Q , s \rangle
\ee 
is then analogous to  the Schema's matching of values, under transforming both states and quantities by the duality maps $d_{\cal S}$ and $d_{\cal Q}$: cf.~Eq.~\eq{obv1}.  (More generally: as in Section \ref{schema}, we take quantum symmetries to also preserve off-diagonal matrix elements:
\bea\label{qsyms}
\bra a(s') |\,\hat Q\, | a(s) \ket=\bra s'|\,\hat Q\,|\,s\ket~,~~~\forall~s,s'\in{\cal S}~,
\eea
for a salient subset of operators in ${\cal Q}$, usually including the Hamiltonian. And this condition can be weakened, to hold only for a salient subset of states in ${\cal S}$.)

This is of course the reason why the Schema confirms the `giant symmetry' analogy. So far---i.e.~before we focus on the two levels, bare theory above and model roots below---there are just two disanalogies between symmetry and duality:\\
\indent (i): Eq.~\eq{symbasic} uses the identity map on quantities, while Eq.~\eq{obv1} uses a duality map $d_{\cal Q}$:  corresponding to the jargon `invariance' vs.~`covariance', and our phrase `preservation or matching' above; \\
\indent (ii): Eq.~\eq{symbasic} typically holds for a salient subset of the quantities, while the duality condition Eq.~\eq{obv1} holds for {\it all} the quantities: this will be illustrated below.

This notion of symmetry is very simple. But suitably adapted and augmented, it will be sufficient for this paper's purposes. In this Section, we make four comments about it. They are regardless of there being a duality; and of the distinction between levels, i.e.~between a bare theory and model roots. So in this Section, we can just think of a theory as a triple of states, quantities and dynamics: $\bra {\cal S}, {\cal Q}, {\cal D} \ket$. These comments will apply, suitably adjusted, to the rest of the paper. The first, third and fourth comments (Sections \ref{dual}, \ref{dynlsymm}, and \ref{sptthies}) are about the notion being adaptable, including to dynamics. The second comment (Section \ref{salientstipul}) is about the idea of a salient set of quantities.  

\subsection{Dual maps}\label{dual}

We said that a symmetry is a map on states that preserves the values of a salient, usually large, set of quantities. Agreed, it is also usual to think of a symmetry as a map on quantities that preserves values on a salient, usually large, set of states: i.e.~for a given state, the value of the argument-quantity equals the value of the image-quantity. Instead of Eq.~\eq{symbasic}, one would write a symmetry as a map $\alpha: {\cal Q} \rightarrow {\cal Q}$ with:  
\be\label{symbasicqties}
\langle \alpha(Q) , s \rangle = \langle Q , s \rangle.
\ee

But there is no conflict here. The two conceptions are related by duality in mathematicians' sense, not ours (cf. Section \ref{dualdef}). That is: one map is the mathematical dual of the other.  Recall that given any map $a: {\cal S} \rightarrow {\cal S}$,  its {\it dual map}  on quantities, $a^*: {\cal Q} \rightarrow {\cal Q}$, is defined by requiring that for any $s \in {\cal S}$ and $Q \in {\cal Q}$: $\bra a^*(Q), s \ket := \bra Q, a(s) \ket$. It follows immediately that  if  $a: {\cal S} \rightarrow {\cal S}$ is a symmetry in our initial sense, i.e.~$a$ respects the structure of $\cal S$, and Eq.~\eq{symbasic} holds, then the dual map on quantities, $a^*: {\cal Q} \rightarrow {\cal Q}$ is a symmetry in the corresponding sense as regards quantities, given by Eq.~\eq{symbasicqties}:
\be\label{dualmapssymmetries}
\langle Q , a(s) \rangle = \langle Q , s \rangle \; \; \Rightarrow \; \; \bra a^*(Q), s \ket := \bra Q, a(s) \ket = \langle Q , s \rangle.
\ee

Besides, while we began with symmetry as a map on states, and conceived symmetries for quantities as dual maps: one could instead equally well start with quantities. For again, one defines dual maps in the same way. Given any map $\alpha: {\cal Q} \rightarrow {\cal Q}$, we say that its dual map on states, ${\alpha}^*: {\cal S} \rightarrow {\cal S}$, is defined by requiring for all arguments: $\bra Q, {\alpha}^*(s) \ket := \bra {\alpha}(Q), s \ket$. One could then define symmetries for quantities by Eq.~\eq{symbasicqties}, and deduce that if $\alpha$ is a symmetry for quantities, its dual map ${\alpha}^*$ is a symmetry for states, i.e.~obeys Eq.~\eq{symbasic}. 

Likewise in the quantum case, with preservation of matrix elements for a salient set of operators. We can define a symmetry $\alpha$ by:
\bea\label{qsalpha}
\bra s'|\,\hat Q\,|s\ket=\bra s'|\,\a(\hat Q)\,|s\ket~,~~~\forall~s,s'\in{\cal S}~.
\eea
Again, this condition can be weakened, to hold only for a salient subset of states in ${\cal S}$. Similar manipulations to the ones in the classical case\footnote{The proof is as follows. Begin with a symmetry $a:{\cal S}\rightarrow{\cal S}$ defined on states, viz.~Eq.~\eq{qsyms}, and rewrite the left-hand side as:  $\bra a(s')|\,\hat Q\,|a(s)\ket=\bra s'|a^\dagger\hat Q\,a|s\ket$. Comparing this with Eq.~\eq{qsalpha}, we see that this induces a symmetry $\a$, defined on quantities. Namely: $\a(\hat Q)=a^\dagger\hat Q\,a$. And the other way around: if the map $\a$ on quantities decomposes into the left- and right-action of some operator $a$, then this induces a symmetry $a:{\cal S}\rightarrow{\cal S}$ on states.} give that the symmetry, defined as a map $\alpha$ on quantities, induces a symmetry, defined as a map $a$ on states, iff the symmetry map $\a$ decomposes in the following way: \bea\label{atoal}
\a(\hat Q)=a^\dagger\,\hat Q~a~.
\eea
Notice that this correspondence between symmetries $\a$ on quantities and symmetries $a$ on states is not one-to-one. For example, if $a$ commutes with a quantity $Q$, we can have a non-trivial symmetry on states that gives rise to a trivial symmetry  on quantities (i.e.~a map $a$ that is non-trivial, while $\a=\mbox{id}$ is trivial).

\subsection{Salient quantities and states: stipulated symmetries}\label{salientstipul}

We said that a symmetry preserves values for `a salient, usually large, set of quantities/states'.  This general formulation deliberately uses the vague word `salient', since it  varies from case to case which quantities/states it is noteworthy to preserve the values of. But it is worth noticing three sorts of consideration that often mould the choice of quantities/states, i.e.~which quantities/states count as salient. The first, which we label {\em stipulated symmetries}, gives a contrast with how we have written so far; the second and third will get a Subsection of their own (Sections \ref{dynlsymm} and \ref{sptthies}).

So far, we have written as if a theory is always given to us with prescribed sets of states and of quantities, so that the set of symmetries is thereby fixed, once some precise meaning of `salient' is fixed. But as we foreshadowed in footnote \ref{Twsyms}: often in physics, we  ``begin our theorizing'' with symmetry principles. That is: we define a theory's sets of states and quantities in order that they carry a representation of a  given abstract symmetry group: spacetime symmetry groups such as the Poincar\'{e} group being a standard example. In such a case, we will say the symmetries are {\em stipulated}---they are part of the definition of the theory. Usually, we think of these symmetries as maps on states: unitary representations of a spacetime symmetry  group on a Hilbert space being a standard quantum example. 

In general, then, a theory $T$ that is formulated as a triple, $\bra{\cal S},{\cal Q},{\cal D}\ket$, is said to have a {\it stipulated symmetry} if it is formulated as having an automorphism of the state-space, $a:{\cal S}\rightarrow{\cal S}$, that preserves some salient subset of the quantities. The stipulated symmetry thus comes with a choice of which quantities  count as salient, so that their values are ``worth'' preserving. This choice is encoded formally in the definition of the triple and its stipulated symmetry, and it of course bears on the theory's interpretation---it moulds the kinds of interpretations that the theory can be given.  

But note that stipulating a symmetry does {\em not}  imply that  every state has its value preserved for {\em every} quantity definable on the state-space. (And {\em mutatis mutandis} if we conceive a symmetry  as a map on quantities; cf.~Section \ref{dual} above.)  For example, one might stipulate rotational symmetry: more precisely, that  in a quantum theory the Hamiltonian is rotationally invariant, so that the Hilbert space carries representations of SO(3).  But this still allows using quantities $Q$ whose expectation values on some states are not invariant under rotations. Thus even with a stipulated symmetry,  there is a  question of selecting the salient quantities.  This point will recur in Section \ref{dynlsymm}.\footnote{For example, Caulton (2015:~p.~156) distinguishes three natural classes, two of which are extreme and trivialising, while the third is the most interesting: (1) symmetries that preserve {\it all} the quantities ${\cal Q}$ of a theory: these are required to preserve everything, which means changing nothing; (2) symmetries that do not preserve any quantities, i.e.~the set of all bijections on the state-space ${\cal S}$; (3) symmetries that lie between these extremes are the interesting cases: especially symmetries that preserve the `set of physical quantities, i.e.~the quantities that, on the basis of their representing the physical properties and relations, register physical differences' (ibid). \label{caulton}} 

\subsection{Dynamical symmetries}\label{dynlsymm}

 We have presented symmetries as maps that preserve the salient quantities' values (and respect the structure of $\cal S$ or $\cal Q$: cf.~Section \ref{dual}). But we have not mentioned time, i.e.~the fact that values change over time. It is indeed very usual to define a symmetry as a map that  `preserves the dynamics'. Taking a symmetry, as usual, as a map on states, this means, roughly: if a sequence of states is possible according to the dynamics, so is the sequence of image-states. 
 
 We can make this precise by using the framework for dynamics, in both Schr\"{o}dinger and Heisenberg pictures, given in Section \ref{values}, Eqs.~\eq{dyns1} and \eq{dyns2}. We will favour the former.  (As noted there, a full discussion would also address the treatment of time in theories not best thought of as triples, e.g.~theories  formulated using partition functions; again, cf.~footnote \ref{nontriple}.)

On Schr\"{o}dinger dynamics, a dynamically possible  total history of the system is a curve through the state-space ${\cal S}$ parameterized by time $t$: with each point $s(t)$ defining the values $\langle Q , s(t) \rangle$ of the various quantities $Q$ at $t$. Then we can define a dynamical symmetry as a map $a$ on $\cal S$ that (a) respects $\cal S$'s structure (b) maps any dynamically possible  history (curve through state-space) to another such history. That is: if $s(t)$ is a dynamically possible  history, the sequence $a(s(t))$ of states is also dynamically possible.\footnote{Preserving the dynamics in this sense is of course a commutation i.e.~equivariance condition. For if we write $s(t) = D_{t, t_0}(s(t_0))$ with $D_{t, t_0}$ representing the  dynamics (cf.~eq \ref{dyns1}),  then preservation of the dynamics is: $a(s(t)) \equiv a(D_{t, t_0}(s(t_0))) = D_{t, t_0}(a(s(t_0)))$.\label{dynsequivariance}} 

On Heisenberg dynamics, the definition of a dynamical symmetry is (we think!) more complicated, because the representation of a dynamically possible history is more complicated. A history is given by a fixed $s \in {\cal S}$ and a family of curves through ${\cal Q}$, all parameterized by time $t$: with $Q_1, Q_2$ on a common curve representing the same physical quantity, e.g.~energy, at two times $t_1, t_2$. So for a single history, there are as many curves through $Q$ in the family as there are physical quantities pertaining to the system. So a dynamical symmetry must be a map whose domain is, not $\cal Q$, but the set of all such families of curves (or all such families that are indeed dynamically possible, once some $s \in {\cal S}$ is chosen). So the map will have to suitably respect, not so much $\cal Q$'s structure, but the structure $\cal Q$ induces on this set of  families of curves. And for the map to be a dynamical symmetry, it must leave invariant the dynamically possible families (allowing, no doubt, for a change of state $s \in {\cal S}$).  But in this paper, we can focus on symmetries as maps on states, and so we will not need to further consider the Heisenberg picture. 

Given this definition of dynamical symmetry in Schr\"{o}dinger picture, as a map on ${\cal S}$ that commutes with the dynamics (footnote \ref{dynsequivariance}), the obvious first question is: how is this related to our initial idea of symmetry as a map on ${\cal S}$ that preserves quantities' values, regardless of time? 

{\em A priori}, they seem very different. Indeed the notion of dynamical symmetry seems weaker in that it  requires the transform $a(s(t))$ of each dynamically possible history $s(t)$ `only' to be itself dynamically possible---it need not have any distinctive relation to $s(t)$, e.g.~by being in some sense a `replica' of $s(t)$. But in fact the notions are drawn together by dynamical symmetry's requirement that the map on histories be induced by a map on states. That is: writing a history as a set of points in ${\cal S}$ for brevity, $\{ s(t) \}$: a dynamical symmetry requires that the map on histories $\{ s(t) \} \mapsto \{ s'(t) \}$ be induced by a map on instantaneous states, i.e.~$s(t) \mapsto a(s(t))$. This turns out to be a strong requirement, thanks to the `sensitivity' of dynamics to the values of many quantities.   That is:  it turns out to force $a$ to preserve the values of many quantities---leading us back to our initial idea of symmetry. But note that this implication is not {\em a priori}: it depends on what dynamical evolution, in typical theories, in fact depends on. 

The point here is well illustrated by spacetime symmetries, as mentioned in Section \ref{salientstipul}. Take for example, spatial translation or spatial rotation in Euclidean space $\mathbb{R}^3$; and consider any of Newtonian mechanics, relativistic mechanics, quantum mechanics, or indeed their `cousin' field theories. Each of these is of course a framework for theorising: not a specific theory, with specific particle and-or field contents, and their dynamics (equations of motion). But it turns out that most such specific theories that have been empirically successful,\footnote{`Successful' within limits, of course: for example, relative velocities small compared with $c$ for Newtonian mechanics to be successful, and typical actions large compared with $h$ for Newtonian or relativistic mechanics to be successful.} once set in Euclidean space $\mathbb{R}^3$, do have spatial translations and spatial rotations as dynamical symmetries. And since a dynamical symmetry is to be induced by a single map $a$ on instantaneous states, e.g.~by spatial translation of 1 mile due East applied to every state, the transform $a(s(t))$ of each dynamically possible history $s(t)$ will indeed be a `replica' of $s(t)$, e.g.~spatially translated by 1 mile. Besides, the dynamics is `sensitive' to the values of many quantities, in the sense that a dynamical symmetry must {\em not} alter them: so its map $a$ on instantaneous states is indeed a symmetry {\em \`{a} la} our initial idea. Again, spatial translations and spatial rotations give a standard illustration; as follows. As we said, most of the empirically successful specific theories written in any of the above frameworks, and set in Euclidean space $\mathbb{R}^3$, have these as dynamical symmetries. But they ``don't allow squeezing''! That is, a dynamical symmetry $a$ must preserve all the relative distances, and relative velocities, between the constituents of the system: if $a$ is, or includes, a spatial translations or rotation, it must be a rigid one, of the system as a whole. It must preserve the values of many quantities---the relative ones.

This discussion of a dynamical symmetry leads in to Section \ref{sptthies}. Also, the cautionary note at the end of Section \ref{salientstipul} above applies again. That is: stipulating that a symmetry be dynamical---indeed, stipulating more specifically: both a symmetry and a dynamics it respects---does not  imply that  every state has its value preserved for  every quantity definable on the state-space. Again, the  example of $\mbox{SO}(3)$ in quantum theory suffices. One might stipulate that $\mbox{SO}(3)$ be a symmetry, and that the Hamiltonian be $\mbox{SO}(3)$-invariant (i.e.~in obvious notation: $[H, U_R] = 0$ for all $R \in \mbox{SO}(3)$). This does not imply that one can only use quantities $Q$ that are rotationally invariant (i.e.~$[Q, U_R] = 0$). 

\subsection{Spacetime theories  and their symmetries}\label{sptthies}

In philosophical and foundational discussions of spacetime theories, it is usual to define symmetries in an apparently different way from ours.

Besides, it is usual to define such theories, not as a triple $\bra {\cal S}, {\cal Q}, {\cal D} \ket$ as we have done, but as a set of, so to speak, possible universes. That is: as a set of $n$-tuples, consisting of a spacetime manifold $M$, equipped with both chrono-geometric structure (encoded in a metric field, a connection etc.) and matter fields (encoded in tensor and spinor fields obeying equations of motion). Each such $n$-tuple represents a (total, 4-dimensional) solution of the theory: a `possible universe'.\footnote{So these $n$-tuples are usually called `dynamically possible models' of the theory: but we will resist yet another use of the over-worked word `model'!} A symmetry is then usually  defined along the following lines. (Recall the example of Galilean  transformations in Section \ref{expleGalil}.) It is a bijection of the manifold that:\\
\indent \indent \indent (a) respects its topological and differential structure (technically: is a diffeomorphism); and whose induced maps on tensor fields, connections etc.:\\
\indent \indent \indent (b) fix the chrono-geometric structure (i.e.~maps the metric field, connection etc.~into themselves) and also \\
\indent \indent \indent (c) send the matter fields into another solution of the equations of motion---another sequence of values over time that is dynamically allowed/possible.

So we need to link our construal of a theory as a triple, and of a symmetry as a map on a state-space, to these ideas. 
The main link is of course that while a physical theory usually has as its subject-matter some limited kind of system, for which we think of the instantaneous state (values of quantities) changing over time,  a spacetime theory takes the universe-throughout-all-time as its subject-matter. So in our construal, a dynamically possible  total history of the system is, on Schr\"{o}dinger picture dynamics, as discussed in Section \ref{dynlsymm}: a curve through ${\cal S}$ parameterized by time $t$, with each point $s(t)$ defining the values $\langle Q , s(t) \rangle$ of the various quantities $Q$ at $t$.\footnote{As  in Section \ref{dynlsymm} , one can have a corresponding discussion using the Heisenberg picture. But here, we set this aside.} But in a spacetime theory, a dynamically allowed/possible total history of the system is just  an $n$-tuple. 

One natural way to  link to our construal is to make a space-vs.-time split within the spacetime theory's manifold.\footnote{An alternative way to link to the usual covariant formulations of spacetime theories takes a model of $T$ to have as its state-space the set of all admissible spacetimes that are solutions of the theory's dynamical equations (e.g.~in general relativity, the Einstein equations). But adopting either of these ways, here and in the main text, one faces issues, if a state is a whole spacetime, about how to ``get inside a spacetime'', so as to distinguish a quantity's differing values  at different spacetime points. Thus one widely adopted approach is to associate quantities not with points, but with extended regions of different types, i.e.~one gets `quasi-local quantities': see Penrose (1988) and Brown and York (1993); for reviews, see Wang (2015) and Szabados (2009).
There are also other issues about which much could be said, such as: rigorously defining quasi-local quantities; fixed fields; the definition of diffeomorphism invariance and its interplay with boundary conditions. (See for example Pooley (2017:~p.~117) and De Haro (2017)).\label{GRcov}}  That is: we take the spacetime theory to have:\\
\indent \indent [a] a state-space $\cal S$ of the instantaneous states of a notional 3-manifold $\Sigma$, which we take as a fiducial spacelike slice of the spacetime manifold; \\
\indent \indent [b] a set of quantities  $\cal Q$  defined on $\Sigma$ (local densities of matter fields etc.); and \\
\indent \indent [c] a dynamics $\cal D$ determining the evolution of the instantaneous state of $\Sigma$. \\
Then a dynamically possible  total history is a foliation of spacetime, whose leaves are time-evolutes of $\Sigma$, equipped with fields. In other words: it is a curve through $\cal S$ parameterized by a time $t$ labelling the leaves of the foliation. 

Combining [a]-[c] with our discussion of dynamical symmetries in Section \ref{dynlsymm}, we can now see that our initial simple idea of a symmetry, as a map $a$ on the system's state-space $\cal S$ that respects its structure and preserves the values of a salient set of quantities, is, after all, similar to the usual definition of symmetry for spacetime theories, (a)-(c) above. For making the space-vs.-time split, [a]-[c], renders this usual definition of symmetry, (a)-(c), like Section \ref{dynlsymm}'s definition of dynamical symmetry. Indeed, most of the familiar (and empirically successful) theories set in the framework of Newtonian mechanics, or relativistic mechanics or quantum mechanics (or their `cousin' field theories) mentioned in Section \ref{dynlsymm} can be written down as spacetime theories, i.e.~as postulating a spacetime manifold equipped with chrono-geometric structure and matter fields: with the different notions of symmetry being linked by using a space-vs.-time split. 

There are of course  several issues here, about which much more could be said. Among them are:\\
\indent \indent (i) the interplay of the structures (topological, differential, metrical etc.) of physical space, spacetime, and state-space; \\
\indent \indent (ii) the justification for singling out, in the spacetime definition of symmetry, the chrono-geometric structure of the manifold as having to be fixed (requirement (b)),  while  the matter fields need not be (especially in the context of general relativity); and \\
\indent  \indent (iii) the justification for making the space-vs.-time split, especially in the context of relativity theory (especially general relativity: cf.~footnote \ref{GRcov}). \\
And of course, such issues are mutually related. For example, (i) and (ii): one might argue that the requirement (b), to fix the chrono-geometric structure, reflects the requirement in our initial idea of symmetry, that $a$ must respect the structure of the state-space $\cal S$, and the fact that $\cal S$'s structure is largely determined by the chrono-geometric structure of spacetime.  However, for this paper we will not need to explore these issues: sufficient unto the day ...

\section{Symmetries of Theories and of Models}\label{symmths}

We now combine the ideas of Section \ref{symmygenl} with Section \ref{schema}'s distinction between a bare theory $T$ and its {\em models} in our sense, viz.~representations of $T$, each with a specific structure of its own. Our notation distinguishes the model's representation of the bare theory's triple, from its specific structure $\bar M$.  It gives a subscript $M$ to the former to signal that it is built out of the latter: $M=\bra{\cal S}_M,{\cal Q}_M,{\cal D}_M;\bar M\ket$. Cf.~Eq.~\eq{eqmodel}. We also wrote this as $M =:\bra m;\bar M\ket~,$ where $m:= T_M := \bra {\cal S}_M, {\cal Q}_M, {\cal D}_M\ket$ is the model triple. 

As we announced in the prospectus (Section \ref{prospectus}), the differing amounts of structure at the two levels, bare theory and model, prompt  some distinctions, and some definitions of types of symmetry. 

We will begin with the idea of symmetries of a theory that are implemented in all the models, which leads to the idea of {\it stipulated symmetries} (Section \ref{secure}). Then we consider symmetries that {\it fail} to be implemented in some of the models, which leads to the idea of {\em accidental symmetries} (Section \ref{fail}). 
We will see that this failure can happen for two diverse reasons. Roughly speaking: either  the theory's homomorphic copy has ``lost the structure with which to exhibit the symmetry'', or  the model has ``extra structure that blocks the symmetry''. 

Then in Section \ref{proper}, we will discuss cases where the specific structure in a model has a symmetry that is  ``invisible'' in the weaker structure of the model triple (the homomorphic copy of the bare theory). Here, `invisible' will mean, roughly speaking: `cannot be defined, or reduces to the trivial symmetry, i.e.~the identity map'. We call such a symmetry a {\em proper symmetry}  of the model. Note that these cases yield a contrast with the other, more common, usage of `model' as an individual solution of a theory. On that usage, a model is  in general {\em less} symmetric than its theory---as is often remarked, with the buzz-word `symmetry-breaking'. A solution of a dynamics with a spherically symmetric Hamiltonian need not be spherically symmetric; a cubical crystal lattice with one particular placing of its lattice points, and one particular orientation of its edges, can be a solution of a dynamics that is translation-invariant and isotropic; and so on. So {\em beware}: our cases of proper symmetries of a model are cases `lying in the opposite direction' from what one expects from the other, more common, usage of `model'. 

Finally, Section \ref{classify} will sum up the Section. It emphasises that these three types of symmetry are mutually exclusive. (It also mentions a fourth type: which will not be important, but which completes our classification, i.e. makes the four types exhaustive.)

This Section builds on Section \ref{symmygenl}.  For in this Section:\\
\indent  \indent (i): we will again take a symmetry as a map $a$ on states, though we could  equally well take it as a map on quantities; \\
\indent \indent (ii): we will return to the idea of stipulated symmetries; and \\
\indent \indent (iii):  one can, at each stage of the discussion, require the symmetry to be a dynamical symmetry; but we will not explicitly mention this; \\
and these three points echo Sections \ref{dual}, \ref{salientstipul} and \ref{dynlsymm}, respectively.

\subsection{Guaranteeing that a bare theory's symmetry is implemented in a model: stipulated symmetries}\label{secure}

For any theory $T$ and any of its models $M$, there is a natural condition for a symmetry of $T$ to be itself realized in $M$: for it to have, so to speak, a ``shadow'' in the model $M$. It is that an obvious diagram should commute (cf.~Figure \ref{commautom}). Here, we write representation as a map  $h$ (`h' for homomorphism). So treating symmetries as maps on states: a symmetry of  $T = \bra {\cal {S}}, {\cal {Q}}, {\cal {D}} \ket$ is a map $a: {\cal S} \rightarrow {\cal S}$ that preserves the value of all quantities in a salient subset of the quantities ${\cal Q}$. So we take $h$ as an appropriate structure-preserving map: from ${\cal S}$ in the theory $T$ itself, to ${\cal S}_M$ in the representing model triple $m = T_M = \bra {\cal S}_M, {\cal Q}_M, {\cal D}_M\ket$. Then the condition---that the symmetry $a$ is itself realized in $M$---is that there should be a map $a_M: {\cal S}_M \rightarrow {\cal S}_M$, such that the diagram in Figure \ref{commautom} commutes.

\begin{figure}
\begin{center}
\bea
\begin{array}{ccc}{\cal S}&\xrightarrow{\makebox[.6cm]{$\sm{$a$}$}}&{\cal S}\\
~~\Big\downarrow {\sm{$h$}}&&~~\Big\downarrow {\sm{$h$}}\\
{\cal S}_M&\xrightarrow{\makebox[.6cm]{$\sm{$a_M$}$}}&{\cal S}_M\nonumber
\end{array}\nonumber
\eea
\caption{Commutativity diagram of the symmetry $a$ with the representation map $h$.}
\label{commautom}
\end{center}
\end{figure}

But this condition is not automatic: even in simple cases, like elementary group theory. In Section \ref{fail} we will discuss how the condition can fail: here we consider ways to secure it. 

There are two obvious ways to do so: i.e.~ways to secure that a bare theory's symmetry is implemented in a  model of it. The first is a formal triviality; the second is a stipulation.

First: consider the case where the model triple is {\em isomorphic} to the bare theory. (In the language of representation theory: the representation is {\em faithful}, i.e.~the homomorphism has trivial kernel.) With the map $h$  an isomorphism, the map $a_M$ 
is bound to exist; and the diagram in Figure \ref{commuteforelmtarygroups1} will trivially commute.
In this case, a symmetry of the theory has a ``duplicate'' or ``replica'' (not just a ``shadow'') in the symmetries of the model---so that in effect, the symmetries of the theory are a {\em subset} of the symmetries of the model. We say `in effect' just because of the different domains of definition: ${\cal S}$ vs.~${\cal S}_M$.

Second: consider, so to speak, the power of stipulation. That is: recall the idea of {\em stipulated symmetries}, as in Section \ref{salientstipul}. The idea there  was that often in physics, we ``begin with'', i.e.~are guided by, symmetry principles, and therefore  {\em require} the state-space  of our theory  to carry a representation of a group, with values of the salient quantities being preserved under the action of the (representation's) group elements. One standard example is requiring the states, of whatever type of particle or field, to carry a representation of a spacetime symmetry group, such as the Poincar\'{e} or Galilean group. 

Carrying this idea over to our two levels, of bare theory and models: it means requiring each model triple to have a symmetry that is a shadow of the bare theory's symmetry. In terms of symmetries as maps on states: each model triple is simply required to have a symmetry $a_M$ on ${\cal S}_M$ that makes Figure \ref{commautom} commute. In short: one has commutation by stipulation.   We shall call such symmetries---both $a$ on the bare theory that `does the requiring', and the `required' $a_M$ on the model $M$---{\em stipulated symmetries}. And similarly for a whole group of symmetries. And again, a spacetime symmetry group gives a standard example. 

This second idea can be implemented in practice as follows: (i) Stipulate the symmetries to be part of the {\it definition} of the bare theory, $T$ (cf.~footnote \ref{Twsyms}). (ii) Reduce the set of models to only those models that instantiate the symmetry, i.e.~those for which the diagram in Figure \ref{commautom} commutes.
Here,  (i) implies (ii).

Thus, for a field theory based on the homogeneous Lorentz group, one can systematically construct the fields by looking at the irreducible representations of this group.\footnote{For textbooks on quantum field theory that take this approach from their very first pages, see Weinberg (1995:~Chapters 2.4-2.7, 5) and Maggiore (2005:~Chapter 2).} Thus we begin (cf.~Weinberg (1995:~Chapters 5.2-5.6)) with the trivial representation of the Lorentz group, which gives a massive scalar field. The next (in terms of its dimension) irreducible representation is the Dirac spinor representation describing a fermion. The next irreducible representation is the (massive) vector representation, i.e.~a four-vector (which contains spin zero and spin one states); and we can have various tensor representations, etc.\\

Often, in practice, and in particular for spacetime symmetry groups: these two ways to secure the commutation---to secure that a bare theory's symmetry is implemented---come together. For the stipulation made is not {\em just} that the state-space  of the model triple carry a symmetry, or carry some or other representation of a group of symmetries. It is also stipulated that it carry a {\em faithful} representation of the group. That is: the representation map $h$ is required to be an isomorphism. So the upshot is: a symmetry of the bare theory has a duplicate or replica (not just a shadow) in the symmetries of {\it each} model (but of course represented differently in the theory and in each of the models)---so that in effect, the stipulated symmetries of the theory are a {\em subset} of the symmetries of the models.\footnote{Again, we say `in effect' just because of the different domains of definition: ${\cal S}$ vs.~${\cal S}_M$. And as mentioned in this Section's preamble: note the contrast with the other, more common, meaning of `model' as an individual solution of a theory. On that meaning, a model is  in general {\em less} symmetric than its theory.}

\subsection{A bare theory's symmetry might fail to be implemented in a model: accidental symmetries}\label{fail}

As we said at the start of Section \ref{secure}, Figure \ref{commautom} need not commute---even in simple cases, like elementary group theory. In fact, the condition can fail in either of two ways: because the theory's homomorphic copy has ``lost the structure with which to exhibit the symmetry''; or because the model has ``extra structure that blocks the symmetry''. We will give an example of each of these: the first from group theory, the second from elementary spacetime theory.

So first, let the bare theory be just a group $G_1$,\footnote{To be a bare theory in our sense, we would need to add a set of maps to the real numbers, to express evaluation of the quantities, and even a dynamics. But to give the simplest case, we ignore these maps.} with an automorphism $a_1: G_1 \rightarrow G_1$; and suppose a group $G_2$ represents $G_1$ thanks to the existence of a homomorphism $h: G_1 \rightarrow G_2$. So  $G_2 \cong G_1/\mbox{ker}\,h$. Then for there to be a homomorphism of $G_2$, $a_2: G_2 \rightarrow G_2$ (even homomorphism: let alone automorphism), that realizes $a_1$ (counts as $a_1$'s ``shadow''  in $G_2$) requires commutation: i.e.~for all $g_1 \in G_1, h(a_1(g_1)) = a_2(h(g_1))$. Or as a diagram, see: Figure \ref{commuteforelmtarygroups1}. But $a_1$ and $h$ need not mesh in this way: commutation of the diagram is not guaranteed. The smallest counterexample we have found is presented in the Appendix. 

\begin{figure}
\begin{center}
\bea
\begin{array}{ccc}G_1&\xrightarrow{\makebox[.6cm]{$\sm{$a_1$}$}}&G_1\\
~~\Big\downarrow {\sm{$h$}}&&~~\Big\downarrow {\sm{$h$}}\\
G_2&\xrightarrow{\makebox[.6cm]{$\sm{$a_2$}$}}&G_2
\end{array}\nonumber
\eea
\caption{Commutativity of group automorphism $a_1$ with group homomorphism $h$.}
\label{commuteforelmtarygroups1}
\end{center}
\end{figure}

A second example of a theory's symmetry failing to exist at the level of its models is given by Section \ref{expleGalil}'s neo-Newtonian spacetime as a bare theory, modelled by various Newtonian spacetimes each with their own specification of absolute rest. Here it is ``extra structure in the models'' that blocks the implementation of the theory's symmetry. Thus we take neo-Newtonian spacetime to be so defined that the Galilean group is its symmetry group: i.e.~the set of diffeomorphisms of the spacetime that fix the spatial metric (giving Euclidean geometry in the instantaneous slices), the temporal 1-form (ordering the slices like $\mathbb{R}$), and the flat affine connection (giving a notion of absolute acceleration, without a notion of absolute rest or absolute velocity). (Cf. Section \ref{sptthies}.) Thanks to avoiding an absolute rest, the boosts are symmetries. But in a Newtonian spacetime, with the extra structure of a specified absolute rest, the boosts are not symmetries: the symmetry group is reduced to the Euclidean group (i.e.~rotations and spatial translations) combined with time translations. In our jargon, whereby a model comprises specific structure as well as a model triple: the specific structure prevents the symmetry. \\

{\it Accidental symmetries:---} We have seen that, in general, the symmetries of a theory are not instantiated in its models. But this of course does not forbid special situations in which a theory happens to have symmetries that are not stipulated (they are not part of the definition of the bare theory). And these symmetries may (though of course they need not) have ``shadows'' in some (but not all) of their models. This is the situation where the diagram in Figure \ref{commautom}, for a specific model $M$, commutes ``for free'', i.e.~without the symmetry $a$ being part of the definition of the bare theory $T$. We call these {\em accidental symmetries}. 

To  illustrate the practical usefulness of the notion of accidental symmetries, we now give a very simple example that we will generalise in the next Section. Consider the bare quantum theory of the hydrogen atom, with a Hamiltonian that only contains a kinetic term for the electron, and a potential that only contains the Coulomb attraction, i.e.~the basic textbook example of the hydrogen atom, neglecting all other interactions. Thus the states of our theory are square-integrable wave-functions, the quantities are the set of operators of the hydrogen atom (in particular the Hamiltonian), and the dynamics is the Schr\"odinger equation (thus we adopt the Schr\"odinger picture). This theory has many models, in our logically weak sense of `model' from Section \ref{model}. Indeed, each square-integrable wave-function that solves the Schr\"odinger equation with given initial conditions (along with the set of operators that act on it, and its time evolution) is a representation of the theory---even if a very non-faithful representation.\footnote{This recovers the usual meaning of `model', (i) in Section \ref{model}, as a {\it special case} of `model', in our sense. Cf.~footnote \ref{usualusage}. \label{usage2}} Let us call these models $M_{n\ell m}$, where $n,\ell,m$ are the quantum numbers of the hydrogen atom.\footnote{Since each state is determined by a set of initial conditions, it is in general not the case that only states with fixed values of $n,\ell,m$ are allowed. The general allowed state is a linear combination of these, depending on the boundary conditions. But for notational simplicity, we label the models by quantum numbers.} But the set of all such models, i.e.~the set of all square-integrable wave-functions that solve the Schr\"odinger equation, is itself also a model. Let us call this model $M_{\sm{total}}:=\{M_{n\ell m}\}$, where $n,\ell,m$ have the usual ranges. 

Let us now discuss the spherical symmetry of this  bare theory of the hydrogen atom. Particular solutions (except for special, spherically symmetric, solutions) are of course not spherically symmetric. Thus, most of the individual models $M_{n\ell m}$ do {\it not} enjoy the SO(3) symmetry of the theory (although they are mapped into each other by the action of SO(3)). On the other hand, $M_{\sm{total}}$ {\it does} carry non-trivial representations of SO(3), which acts as an automorphism of the set of states. Namely, SO(3) maps states with the same $n$ and $\ell$ but different values of $m$, so that $\ell$ labels the distinct representations of SO(3) (and, by focussing on a suitable set of eigenfunctions of the Laplacian that form a complete set, one can get the irreducible representations). In other words, SO(3) is a symmetry of $M_{\sm{total}}$. Of course, all of the above   generalises to the case where the potential  is specified  only to be a spherically symmetric function, i.e.~of the form $V(r)$, and we consider other spatial dimensions: as we will do in the next Section.

Thus rotational symmetry is an accidental symmetry of this particular theory. It is a symmetry of the theory that is represented non-trivially in one of the models of the theory, but not in all the models. 

\subsection{Proper symmetries of models}\label{proper}

In the last two Sections, we considered symmetries of a theory that are stipulated to be symmetries of all the models (Section \ref{secure}) or that fail to be realized in at least some of the models (Section \ref{fail}). In this Section, we consider symmetries that are {\it not} defined in the theory, but that are nevertheless symmetries of a model.
For a model's specific structure $\bar M$---its `content' that goes beyond its being a model/realization of $T$---can make the model have symmetries  {\it additional} to those that are ``shadows'', or even ``duplicates'', of the symmetries of $T$. And we expect that if these additional symmetries are well-defined on the model triple, or if they naturally induce a symmetry there, that symmetry is trivial, i.e.~just the identity map on the model triple. 

Our prototypical cases of representations of a group or algebra give examples. Perhaps the simplest is as follows. Let $T$ be the real numbers $\mathbb{R}$; and let $M$ be the complex numbers $\mathbb{C}$ which of course represents $\mathbb{R}$ as the real axis, i.e.~the complex numbers with zero imaginary part,  $\{ z \in \mathbb{C} \, | \, z = x + i0, \, x \in \mathbb{R} \}$. So this latter set, the real axis, is like the model triple. Then $M$ has the symmetry of complex conjugation $z \mapsto {\bar z}$ which is indeed well-defined on the real axis: but there, it is just the identity map.\footnote{There are also examples in interesting cases of dualities. In gauge-gravity dualities, De Haro (2017) showed that a certain subgroup of the diffeomorphism group of the gravity model of the theory (roughly, the diffeomorphisms which preserve the asymptotic boundary conditions) was `invisible' to the gauge model of the theory, in the sense of not representing any difference on that model: and so these diffeomorphisms are not in the common core between the two models, and they are trivially represented on the bare theory. The same verdict was made in De Haro (2016a:~\S2.2.3) for the `gauge symmetries' of the gauge side of the duality. These are not visible on the gravity side: they are symmetries of the formulation of the gauge model of the theory, and are trivially represented on the bare theory.} 

This prompts the idea of a {\em proper symmetry} of a model: where the word `proper' is to connote `specific to the model'. The idea is that such a symmetry depends, at least in part, on the model's specific structure---which, recall, ``lies beyond'' the model triple's representing the bare theory: just like complex conjugation in the example just given.\\

To give a general definition, it is clearest to introduce another notation for a model, as a triple of its own set of states, quantities and dynamics. So we write: $ M=\bra{\cal {\bar S}},{\cal {\bar Q}},{\cal {\bar D}} \ket$. Contrast the way our previous notation in Eq.~\eq{eqmodel} separated out the specific structure $\bar M$. Thus we wrote: $ M=\bra{\cal S}_M,{\cal Q}_M,{\cal D}_M ; \bar M\ket  = \bra m ; \bar M\ket$. With the new notation, we can discuss `symmetry of a model' in just the same way that, already in Section \ref{symmygenl}, we discussed `symmetry of a theory': namely, as a map $a$ on the state-space ${\cal {\bar S}}$ that preserves the values of a  salient, usually large, set of quantities. 

This suggests the following definition. Given a (bare) theory $T$ that is represented by various models and presented as a triple, we will say that a symmetry $a$ of one such model $M = \bra{\cal {\bar S}},{\cal {\bar Q}},{\cal {\bar D}} \ket$, $a: {\cal {\bar S}} \rightarrow {\cal {\bar S}}$, is a {\em proper symmetry} of $M$, if there is some other model of $T$, say $M' = \bra{\cal {\bar S'}},{\cal {\bar Q'}},{\cal {\bar D'}} \ket$ for which, in effect, $a$ cannot be defined; and $a$ can, in effect, also not be defined for the theory $T$. (Here, we say `in effect' for the same reason as in Section \ref{secure}: i.e.~just because of the different domains of definition: $\cal {\bar S}$ vs.~$\cal {\bar S'}$.) More precisely: in the model $M' = \bra{\cal {\bar S'}},{\cal {\bar Q'}},{\cal {\bar D'}} \ket$, there is no natural definition of a symmetry  map $a': {\cal {\bar S'}} \rightarrow {\cal {\bar S'}}$ that is a ``cousin'' of $a$, except perhaps as the identity map.

Again, the complex numbers vs.~the real numbers give what is perhaps the simplest example, with the real line now taken as another model $M' := \mathbb{R}$, along with the  complex plane $M := \mathbb{C}$: rather than $\mathbb{R}$ being the theory $T$. 
In this example, complex conjugation is a proper symmetry of $M := \mathbb{C}$: it reduces to the identity map on the other model, $M' := \mathbb{R}$.\\

Consider again the example from the previous Section, but in generalised form. We stipulate that our bare theory is the quantum mechanics of the abstract two-body problem in an arbitrary number of spatial dimensions. Consider now models that specify:\\
\indent (i)~the number of spatial dimensions, $n$, and the idealization that one body is a test-body, which secures rotational invariance, and so an $\mbox{SO}(n)$ symmetry; and \\
\indent (ii)~the form of the potential. 

Specifically, we will take the model to specify the dimension of space to be 3, and the potential to be the Coulomb potential. With these definitions, the SO(3) rotational symmetry is not a symmetry of the bare theory but it is a symmetry of the model. Furthermore, the model's specifying the potential to decay like $1/r$ means that the SO(3) symmetry is enlarged to SO(4), where the additional symmetry corresponds to a new conserved quantity---the Runge-Lentz vector.\footnote{Recall that the rotational invariance of the Hamiltonian implies the conservation of angular momentum, which means that the classical orbits lie on a plane. Likewise, the SO(4) symmetry implies the conservation of the Runge-Lenz vector, namely ${\bf M}:=({\bf p}\times{\bf L}-{\bf L}\times{\bf p})/2m-e^2{\bf r}/r$: i.e.~it commutes with the Hamiltonian.} On the other hand, notice that there are countless other choices of potential that do not share the SO(4) symmetry. Thus, in fact, this model has an SO(4) symmetry that the bare theory and countless other models do not share: it is a proper symmetry of this model.

\subsection{Kinds of symmetry}\label{classify}

Let us sum up this Section's discussion of how the symmetries of a bare theory relate to those of its models. We have made two main points, as follows.\\
\indent \indent (i):    A bare theory $T$ is represented by one of its model triples, $m$. The model $M$ then consists of  $m$ and some specific structure $\bar M$; (cf.~Section \ref{jargonnot}). But representation requires only a homomorphism, not an isomorphism. Hence our articulating in Sections \ref{secure} and \ref{fail} the condition---in terms of a commuting diagram---for a symmetry of $T$ to be itself realized in $m$. As Section \ref{fail} described, this condition can fail. So when this condition {\it is} met for some of the models---not by stipulation but ``accidentally''---we say that there is an {\it accidental symmetry}. \\
\indent \indent (ii): When the diagram is stipulated to commute, so that each model has a `duplicate' or `replica' of the symmetry $a$, we have a {\it stipulated symmetry}. There are two obvious ways to implement this stipulation: defining the theory to have a symmetry, and accordingly restricting the set of models to those that respect the symmetry, or requiring the model triples to be isomorphic to the theory $T$ (the latter condition being strictly stronger). But even in such cases, the {\em model} (as against the model triple) has its own specific structure $\bar M$---which may have symmetries that $m$, and the other models ``know nothing of'': these are {\it proper symmetries of the model} (Section \ref{proper}). 

We can also now see that our three types of symmetry are mutually exclusive. Stipulated symmetries and accidental symmetries are mutually exclusive because the former are stipulated while the latter are not; and stipulated symmetries and proper symmetries of models are mutually exclusive because, while the former has a shadow (i.e.~can be defined as a symmetry) in all of the models, the latter cannot be defined in at least some models (and it is also not defined in the theory).

Accidental symmetries and proper symmetries are mutually exclusive because an accidental symmetry is a symmetry of the theory (although it is not stipulated); while a symmetry of the theory cannot be a proper symmetry of a model.

The only possibility that is not covered by the above three kinds of symmetries is the possibility that a symmetry is a symmetry of all the models but is not a symmetry of the theory. Such a symmetry can for example be obtained when a proper symmetry of the models is shared by some models (but of course not all), and then the models for which this symmetry is not realized are excised from the set of models. We could call such a symmetry an {\it improper symmetry of the models} (improper, in the sense that it is a symmetry of all the models but not of the theory). 

Indeed, it is now true that any symmetry (in our sense) belongs to one of these four categories. For, given a theory and a set of models, a symmetry can either be a symmetry of the theory and of all the models (hence it is stipulated); or it is a symmetry of all the models but not the theory (an improper symmetry); or it is a symmetry of the theory and of only some models (an accidental symmetry); or it is a symmetry of some models but not of the theory and not of some other models (a proper symmetry of the models); or it is only a symmetry of the theory (also an accidental symmetry).

\section{Duality Preserves Symmetries}\label{reltosymmy}

It is straightforward to confirm that on Section \ref{dualdef}'s definition of duality, a duality preserves any symmetry (including dynamical symmetries) of its model triples. There are two points here. First: there is a commuting square diagram of isomorphisms. Second: there is the issue of the values of a quantity being equal on a given state, and on its transform under a symmetry. The first point will lead in to the second. \\

First: The duality maps $d_{\cal S}, d_{\cal Q}$ are not only bijections, but isomorphisms: $d_{\cal S}: {\cal S}_{M_1} \rightarrow {\cal S}_{M_2}$, and $d_{\cal Q}: {\cal Q}_{M_1} \rightarrow {\cal Q}_{M_2}$. And although we did not have to spell out the exact structures of ${\cal S}_{M_i}, {\cal Q}_{M_i}$ that these isomorphisms are to preserve,
it is obvious from the fact that `is isomorphic to' is both a symmetric and a transitive relation, that the following diagram, with $a$ understood to be any automorphism of  ${\cal S}_{M_1}$, commutes (cf.~Figure \ref{dlykeepsymmystates}).
\begin{figure}
\begin{center}
\bea
\begin{array}{ccc}{\cal S}_{M_1}&\xrightarrow{\makebox[.6cm]{$\sm{$a$}$}}&{\cal S}_{M_1}\\
~~\Big\downarrow {\sm{$d_{\cal S}$}}&&~~\Big\downarrow {\sm{$d_{\cal S}$}}\\
{\cal S}_{M_2}&\xrightarrow{\makebox[.6cm]{}}&{\cal S}_{M_2}
\end{array}\nonumber
\eea\caption{Commutativity of duality and symmetry for states.}
\label{dlykeepsymmystates}
\end{center}
\end{figure}

And  of course, this diagram of isomorphisms is just what we mean by saying a duality $d$ preserves an automorphism of the state-space ${\cal S}_{M_1}$ in its domain model triple, and preserves ${\cal S}_{M_1}$'s structure. Namely, $d$ carries the automorphism---a map $a$  on ${\cal S}_{M_1}$---to a corresponding automorphism of states in the codomain (indeed: range) model triple. The diagram defines this corresponding automorphism, i.e.~the map forming the fourth side of the square: $d_{\cal S} \circ a \circ (d_{\cal S})^{-1}: {\cal S}_{M_2} \rightarrow {\cal S}_{M_2}$. 

There is obviously a corresponding point about quantities, as against states. Since $d_{\cal Q}$ is required to be an isomorphism of quantities, the following diagram, with $a$ now understood to be any automorphism of  ${\cal Q}_{M_1}$, must commute, cf.~Figure \ref{dlykeepsymmyqties}.
\begin{figure}
\begin{center}
\bea
\begin{array}{ccc}{\cal Q}_{M_1}&\xrightarrow{\makebox[.6cm]{$\sm{$a$}$}}&{\cal Q}_{M_1}\\
~~\Big\downarrow {\sm{$d_{\cal Q}$}}&&~~\Big\downarrow {\sm{$d_{\cal Q}$}}\\
{\cal Q}_{M_2}&\xrightarrow{\makebox[.6cm]{}}&{\cal Q}_{M_2}
\end{array}\nonumber
\eea\caption{Commutativity of duality and symmetry for quantities.}
\label{dlykeepsymmyqties}
\end{center}
\end{figure}

And again, this diagram is just what we mean by saying a duality $d$ preserves an automorphism of the quantities in its domain model triple, and preserves ${\cal Q}_{M_1}$'s structure. Namely, $d$ carries the automorphism---a map $a$  on  ${\cal Q}_{M_1}$---to a corresponding automorphism of quantities in the codomain (indeed: range) model triple. The diagram defines this corresponding automorphism: $d_{\cal Q}~ \circ~ a ~\circ ~(d_{\cal Q})^{-1}: {\cal Q}_{M_2} \rightarrow {\cal Q}_{M_2}$. 

In short: Figures \ref{dlykeepsymmystates} and \ref{dlykeepsymmyqties} show that duality commutes with automorphisms of the states and of the quantities.\\

Second: But in physics, the notion of symmetry of course involves more than the notions of automorphism of the state-space, and of the set (usually algebra) of quantities. It involves the pairing whereby states $s$ and quantities $Q$ assign each other a value: $\bra Q, s \ket$. For these values (for a large and salient set of quantities)  must be preserved under the symmetry. 

But satisfying this is automatic, for a duality as defined in Section \ref{dualdef}. That is: For a duality to respect this aspect of symmetry was already built in to our definition of duality: namely in condition (i), that the values are equal between states and quantities that correspond by the duality. Recall Eq.~\eq{obv1}), which we here repeat:
\be\label{obv1repeat}
\langle Q_1, s_1 \rangle_1 = \langle d_{\cal Q}(Q_1), d_{\cal S}(s_1) \rangle_2 \; , \;\; \forall Q_1 \in {{\cal Q}_{M_1}}, s_1 \in {{\cal S}_{M_1}}. 
\ee

Thus let us show that the map $a_2 := d_{\cal S} \circ a \circ (d_{\cal S})^{-1}: {\cal S}_{M_2} \rightarrow {\cal S}_{M_2}$ is a symmetry on states in our sense, i.e.~preserves values; cf. Figure \ref{dlykeepsymmystates}. (A similar argument works for the symmetry on quantities, as in Figure \ref{dlykeepsymmyqties}.) In effect, we just use the duality maps and Eq.~\eq{obv1repeat} to carry back the discussion of preservation of values to the automorphism $a: {\cal S}_{M_1} \rightarrow {\cal S}_{M_1}$ in model $M_1$. Thus we consider $\langle Q_2, s_2 \rangle_2$, and we write $Q_1, s_1$ for the inverse images, under $d_{\cal Q}$ and $d_{\cal S}$ respectively,  of  $Q_2, s_2$. Then we  have (using Eq.~\eq{obv1repeat} for the second and fourth equations, and $a$ being a symmetry for the third equation):
\bea\label{dualpreservevaluesofsymmy}
\langle Q_2, s_2 \rangle_2 &=&  \langle d_{\cal Q}(Q_1), d_{\cal S}(s_1) \rangle_2 = \langle Q_1, s_1 \rangle_1 = \langle Q_1, a(s_1) \rangle_1 \\
 &=& \langle d_{\cal Q}(Q_1), d_{\cal S}(a(s_1)) \rangle_2 = \langle Q_2, a_2(d_{\cal S}(s_1)) \rangle_2 \equiv \langle Q_2, a_2(s_2) \rangle_2 \; .\nonumber
\eea

Finally: the same verdict---that a duality preserves any symmetry of its model triples---applies to dynamics, i.e.~to dynamical symmetries. Recall from footnote \ref{dynsequivariance} (in Section \ref{dynlsymm}) that a dynamical symmetry is a commutation i.e.~equivariance condition. So for the Schr\"{o}dinger picture of dynamics, the diagram for the `first'  side of a duality, i.e.~$m_1 = \bra  {\cal S}_{M_1}, {\cal Q}_{M_1}, {\cal D}_{M_1} \ket$, is, with $a$ the dynamical symmetry, as in Figure \ref{dynslsymmyfordltysection}.
\begin{figure}
\begin{center}
\bea
\begin{array}{ccc}{\cal S}_{M_1}&\xrightarrow{\makebox[.6cm]{$\sm{$a$}$}}&{\cal S}_{M_1}\\
~~\Big\downarrow {\sm{$D_{t,t_0}$}}&&~~\Big\downarrow {\sm{$D_{t,t_0}$}}\\
{\cal S}_{M_1}&\xrightarrow{\makebox[.6cm]{$\sm{$a$}$}}&{\cal S}_{M_1}
\end{array}\nonumber
\eea
\caption{Commutativity of symmetry and dynamics.}
\label{dynslsymmyfordltysection}
\end{center}
\end{figure}

So we now compose this diagram with Figure \ref{dlykeepsymmystates}, which represents that a duality preserves a symmetry. But since in Figure \ref{dynslsymmyfordltysection}, the `first'  side, `1', of the duality occurs twice, on both top and bottom rows, we now need to compose  Figure \ref{dynslsymmyfordltysection} with Figure \ref{dlykeepsymmystates} twice: both on its bottom row; and also  on its top row (with the duality arrow in Figure \ref{dlykeepsymmystates} reversed). The resulting diagram (Figure \ref{threebox}) shows that the duality isomorphism on state-spaces $d_{\cal S}$ carries the dynamical symmetry $a$ on the `1'  side of the duality, to a dynamical symmetry on the `2'  side: namely, the symmetry $d_{\cal S} \circ a \circ d^{-1}_{\cal S}$ (cf.~either the top or bottom square). The Schr\"{o}dinger picture dynamics on ${\cal S}_{M_2}$ is (reading down the columns in the Figure): $d_{\cal S} \circ D_{t,t_0} \circ d_{\cal S}^{-1}$.
\begin{figure}
\begin{center}
\bea
\begin{array}{ccc}{\cal S}_{M_2}&\longrightarrow&{\cal S}_{M_2}\\
~~\Big\downarrow {\sm{$d_{\cal S}^{-1}~~$}}&&~~\Big\downarrow {\sm{$d_{\cal S}^{-1}~~$}}\\
{\cal S}_{M_1}&\xrightarrow{\makebox[.6cm]{$\sm{$a$}$}}&{\cal S}_{M_1}\\
~~\Big\downarrow {\sm{$D_{t,t_0}$}}&&~~\Big\downarrow {\sm{$D_{t,t_0}$}}\\
{\cal S}_{M_1}&\xrightarrow{\makebox[.6cm]{$\sm{$a$}$}}&{\cal S}_{M_1}\\
~~\Big\downarrow {\sm{$d_{\cal S}~~~~$}}&&~~\Big\downarrow {\sm{$d_{\cal S}~~~~$}}\\
{\cal S}_{M_2}&\longrightarrow&{\cal S}_{M_2}
\end{array}\nonumber
\eea
\caption{Commutativity of duality, symmetry, and dynamics.}
\label{threebox}
\end{center}
\end{figure} \\

To sum up: we have shown that a duality always preserves a symmetry of its model triples.\\

\section{Conclusion}\label{conclude}

In this paper we have investigated the relations between dualities and symmetries, using our Schema for dualities. The Schema begins with a conception of theories and of models that is widespread in philosophy of science (namely, a theory as a formal structure or a set of axioms, and models as instantiations or representations of the theory) but that departs from the common usage, in philosophy of physics, of models as individual solutions of the theory's dynamical equations. The flexibility in our usage allows both a natural definition of a duality as an isomorphism between models {\it and} the recovery, as a special case, of the more familiar notion of `model' in philosophy of physics: by appropriately tuning the level of the description---more precisely, by carefully selecting the relevant representations.

In Section \ref{jargonnot}, we emphasised (i) the distinction within a model (in our sense, viz. representation of a bare theory), between model root and specific structure; and (ii) the distinction between internal and external interpretations. This gave us a clear formulation of disagreements like those between Newton (Clarke) and Leibniz, about the status of absolute rest: which we illustrated with Galilean transformations. We also used matrix representations of a group to illustrate the different ways one can define `model root' and `specific structure'. 

The distinction (i) also underpins the Schema's exact definition of duality as an isomorphism of model roots (Section \ref{schema}). This equipped us, after reviewing symmetries in general (Section \ref{symmygenl}), to understand in detail how dualities and symmetries relate to one another. 

This we undertook in the last two Sections. In Section \ref{symmths}, the Schema's conceptions of theory and model prompted us to introduce four types of symmetries: of which two are symmetries of the theory (stipulated symmetries and accidental symmetries) and two are symmetries of models only (proper and improper symmetries). These four types are mutually exclusive and jointly exhaustive of all the possibilities for symmetries, given the Schema's distinction between theory and models, and our construal of `symmetry'. Then in Section \ref{reltosymmy}
we showed that, just as one would hope, the Schema's construal of duality meshes with whatever symmetries the duals, i.e. the isomorphic model roots, may have.

\section*{Acknowledgements}

We thank the organizers and audience at the 2017 conference in Hannover; audiences at conferences in Bad Honnef and London in 2018; and Niels Martens, James Read and two anonymous reviewers for comments. SDH's work was supported by the Tarner scholarship in Philosophy of Science and History of Ideas, held at Trinity College, Cambridge.

\section*{Appendix. A Group Automorphism Lost under Homomorphism}
\addcontentsline{toc}{section}{Appendix. A Group Automorphism Lost under Homomorphism}\label{appendix}

We present a simple case of Section \ref{fail}'s topic, of a symmetry of a theory being ``lost'' in a model. As discussed in  Section \ref{fail}: for the simple case of group theory, this is a matter of the diagram in Figure \ref{commuteforelmtarygroups1} not commuting. That is: an automorphism $a_1: G_1 \rightarrow G_1$, and a homomorphism $h: G_1 \rightarrow G_2 \cong G_1/\mbox{ker}\,h$ need not mesh as the diagram shows, so as to give a well-defined homomorphism $a_2: G_2 \rightarrow G_2$.

To see what is involved, we can identify $G_2$ with its isomorph $G_1/\mbox{ker}\,h$; and let us write $K$ as short for $\mbox{ker}\,h$. Then the commutation requirement $h(a_1(g_1)) = a_2(h(g_1))$ is the requirement that the definition of $a_2$ by saying $a_2(K g_1) := h (a_1(g_1))$ be well-defined. That is: the given automorphism $a_1$ must respect the cosets of $h$. That is: we require that if $h(g_1) = h(g'_1)$, i.e.~$Kg_1 = Kg'_1$, i.e.~there is a $k \in K$ such that $g'_1 = k g_1$, then also there is a $k \in K$ such that $a_1(g'_1) = k a_1(g_1)$. But $a_1$ and $h$ need not mesh in this way: commutation of the diagram is not guaranteed. 

The smallest counterexample we have found takes $G_1$ as the dihedral group $D_4$. It has $D_2$ as a normal subgroup, $D_2 \lhd D_4$, with $D_4 / D_2 \cong C_2 = \{0,1\}$. So we take $G_2$ as (up to isomorphism) the cyclic group $C_2$ and $h: G_1 \rightarrow G_2$ as the canonical projection taking each element of $D_4$ to its $D_2$-coset. Then we define $a_1: D_4 \rightarrow D_4$ to take an element $b$ of the normal subgroup $D_2$ {\em out} of $D_2$. This implies that the diagram cannot commute. For on the left of the diagram $h: D_4 \ni b \mapsto D_2 = e_{D_4 / D_2}$, so that any bottom-row homomorphism $a_2$ must map $D_2 = e_{D_4 / D_2} \mapsto D_2 = e_{D_4 / D_2}$; while on the right of the diagram $h: D_4 \ni a_1(b) \mapsto (D_2)a_1(b) \neq e_{D_4 / D_2}$.  

The details are as follows. $D_4$ is the symmetry group of the square, and is generated by two elements: rotation out of the plane of the square by $\pi$ about the axis in the plane that  horizontally bisects the square; and rotation in the plane of the square by $\pi/2$ about the axis normal to the plane through the centre of the square. We label these $b$ and $c$ respectively. So with $e$ the identity  transformation, $b^2 = c^4 = e$. One checks that $bc$ and $bc^3$ are the rotations out of the plane of the square by $\pi$ about the two diagonal axes, and $bc^2$ is the rotation about the axis in the plane that  vertically bisects the square. So these elements are of order 2. Also one checks that $cb = bc^3, c^2b = bc^2, c^3b = bc$. And so $D_4$ has eight elements, which we write as $\{e, c, c^2, c^3, b, bc, bc^2, bc^3 \}$, with generating equations $b^2 = c^4 = (bc)^2 = e$. $D_4$ has a copy of the dihedral group $D_2$---viz.~$\{e, c^2, b, bc^2\}$ with generating equations $b^2 = (c^2)^2 = (bc^2)^2 = e$---as a normal subgroup: $D_2 \lhd D_4$, with $D_4 / D_2 \cong C_2$. So define $h: D_4 \rightarrow D_4 / D_2 \equiv \{ D_2, D_2(bc) \}$ as the canonical projection. That is: each element of $D_2$ is sent to $D_2 \in D_4 / D_2$; and the other four elements, $c, c^3, bc, bc^3$, are each sent to their common coset $D_2(bc)$. Finally, we define the automorphism  $a_1: D_4 \rightarrow D_4$ as:\\
\indent \indent  (i) the identity map on the rotations in the plane (the powers of $c$); but also\\
\indent \indent (ii) mapping $b$ to $bc$, i.e.~$b \in D_2$ is  mapped out of $D_2$, and \\
\indent \indent  (iii)  mapping $bc$ to $bc^2$, $bc^2$ to $bc^3$, and $bc^3$ to $bc^4 \equiv b$ (so that an element out of $D_2$ is mapped into $D_2$).\\
A tedious check shows that $a_1$ thus defined is an automorphism. So by the argument in the preceding paragraph,  the diagram in Figure \ref{commuteforelmtarygroups1} does not commute. 
 
\section*{References} 

Auslander, L. and MacKenzie, R.~(1963). {\em Introduction to Differentiable Manifolds}. McGraw-Hill: Dover reprint 2007.\\
\\
Bargmann, V.~(1954). `On Unitary Ray Representations of Continuous Groups'. {\it Annals of Mathematics}, 59 (1), pp.~1-46.\\
\\
Belot, G.~(2000). `Geometry and motion'. {\em British Journal for the Philosophy of Science} 51, pp.~561-595.\\
\\
Brading, K.~and Castellani, E.~(Eds.) (2003). {\it Symmetries in Physics: Philosophical Reflections}. Cambridge University Press.\\
\\
Brown, J.~D.~and York, J.~W.~Jr.~(1993). `Quasilocal energy and conserved charges derived from the gravitational action'. {\it Physical Review D}, 47 (4), pp.~1407-1419.\\
\\
Butterfield, J.~(2006). `On symplectic reduction in classical mechanics'. In: J. Earman and J. Butterfield (eds.) {\em The Handbook of  Philosophy of Physics}, North Holland 2006; pp.~1-131. http://arxiv.org/abs/physics/0507194  and http://philsci-archive.pitt.edu/2373/ \\
\\
Butterfield, J.~(2014). `Reduction, Emergence and Renormalization', {\em The Journal of Philosophy} 111, pp.~5-49: http://arxiv.org/abs/1406.4354; http://philsci-archive.pitt.edu/10762/. \\
\\
Butterfield, J.~(2018). `On Dualities and Equivalences Between Physical Theories'. Forthcoming in {\it Space and Time after Quantum Gravity}, Huggett, N.~and W\"uthrich, C.~(Eds.).\\
\\
Callender, C. and Cohen, J. (2006). `There is no special problem about scientific representation'. {\em Theoria}, 21, pp.~67-85; http://philsci-archive.pitt.edu/2177. \\ 
\\
Castellani, E.~(2017). `Duality and particle democracy'. {\it Studies in History and Philosophy of Modern Physics}, pp.~100-108.
  doi:10.1016/j.shpsb.2016.03.002\\
\\
Caulton, A.~(2015). `The Role of Symmetry in the Interpretation of Physical Theories'. {\it Studies in History and Philosophy of Modern Physics}, 52, pp.~153-162.\\
\\
De Haro, S.~(2015). `Dualities and emergent gravity: Gauge/gravity duality'. {\em Studies in History and Philosophy of Modern Physics}, 59, 2017, pp.~109-125. \\doi:~10.1016/j.shpsb.2015.08.004. PhilSci 11666.\\
\\
De Haro, S.~(2016). `Spacetime and Physical Equivalence'. Forthcoming in {\it Space and Time after Quantum Gravity}, Huggett, N. and W\"uthrich, C.~(Eds.),\\ http://philsci-archive.pitt.edu/13243.\\
\\
De Haro, S.~(2016a). `Duality and Physical Equivalence'. \\
 http://philsci-archive.pitt.edu/id/eprint/12279 (This is an expansion of (2016); and the title has changed). \\
\\
De Haro, S.~(2017). `The Invisibility of Diffeomorphisms'. {\it Foundations of Physics}, 47 (11), pp.~1464-1497.\\
\\
De Haro, S.~(2018). `The Heuristic Function of Duality'. {\em Synthese}; https://doi.org/10.1007/s11229-018-1708-9. arXiv:1801.09095 [physics.hist-ph].\\
\\
De Haro, S.~(2018a). `Theoretical Equivalence and Duality'. This volume.\\
\\
De Haro, S.~and Butterfield, J.N.~(2018). `A Schema for Duality, Illustrated by Bosonization'. {\em Foundations of Mathematics and Physics One Century after Hilbert}. J.~Kouneiher (Ed.). 
Springer. \\
\\
De Haro, S., Mayerson, D., Butterfield, J.~N.~(2016). `Conceptual Aspects of Gauge/Gravity Duality'. {\it Foundations of Physics}, 46 (11), pp.~1381-1425. doi:~10.1007/s10701-016-0037-4. 
[arXiv:1509.09231 [physics.hist-ph]].\\
\\
De Haro, S., Teh, N., Butterfield, J.N.~(2017). `Comparing Dualities and Gauge Symmetries'. {\em Studies in History and Philosophy of Modern Physics}, 59, pp.~68-80. [arXiv:1603.08334 [physics.hist-ph]]. http://philsci-archive.pitt.edu/12009\\
\\
De Haro, S.~and De Regt, H.~W.~(2018). `Interpreting theories without a Spacetime'. {\it European Journal for Philosophy of Science}, 8 (3), pp.~631-670. https://doi.org/10.1007/s13194-018-0204-x.\\
\\
De Haro, S.~and De Regt, H.~W.~(2018a). `A Precipice Below Which Lies Absurdity? Theories without a spacetime and scientific understanding'. {\it Synthese}. https://doi.org/10.1007/s11229-018-1874-9.\\
\\
Dewar, N.~(2015). `Symmetries and the philosophy of language'.  {\em Studies in History and Philosophy of Modern Physics}, {\bf 52}  pp.~317-327. \\
\\
Dewar, N.~(2017). `Sophistication about symmetries'. {\it The British Journal for the Philosophy of Science,} 70 (2), pp.~485-521.\\
\\
Dieks, D., Dongen, J. van, Haro, S. de~(2015), `Emergence in Holographic Scenarios for Gravity'. 
{\it Studies in History and Philosophy of Modern Physics}, 52, pp.~203-216. doi:~10.1016/j.shpsb.2015.07.007.\\
\\
Earman, J.~(1989). {\em World Enough and Spacetime}, MIT Press.\\
\\
Fine, A.~(1984). `The Natural Ontological Attitude'. In: {\it Scientific Realism}, J.~Leplin (Ed.), pp.~83-107. University of California Press.\\
\\
Fine, A.~(1986). {\it The Shaky Game}, University of Chicago Press.\\
\\
Fraser, D.~(2017). `Formal and physical equivalence in two cases in contemporary quantum physics'. {\it Studies in History and Philosophy of Modern Physics}, 59, pp.~30-43. \\
\\
Glymour, C.~(2013). `Theoretical Equivalence and the Semantic View of Theories'. {\it Philosophy of Science}, 80, pp.~286-297.\\
\\
Holm, D.~D.~(2011). {\it Geometric Mechanics. Part II: Rotating, Translating and Rolling}. Imperial College Press, 2nd Edition.\\
\\
Huggett, N.~(2017). `Target space $\neq$ space'. {\em Studies in History and Philosophy of Modern Physics}, 59, 81-88. doi:10.1016/j.shpsb.2015.08.007.\\
\\
Kaiser, D.~(2005). {\em Drawing Theories Apart: The Dispersion of Feynman Diagrams in Postwar Physics} Chicago: University Press.\\
\\
Lewis, D.~(1970). `General semantics'. {\em Synthese} {\bf 22}, pp.~18-67; reprinted in his {\em Philosophical Papers: volume 1}  (1983), Oxford University Press.\\
\\
Lewis, D.~(1975). `Languages and Language'. In Keith Gunderson (ed.), {\em Minnesota Studies in the Philosophy of Science, Volume VII}, Minneapolis: University of Minnesota Press, pp.~3-35.\\
\\
Lutz, S.~(2017). `What Was the Syntax-Semantics Debate in the Philosophy of Science About?' {\it Philosophy and Phenomenological Research}, XCV (2), pp.~319-352.\\
\\
Maggiore, M.~(2005). {\it A Modern Introduction to Quantum Field Theory}. Oxford: Oxford University Press.\\
\\
Matsubara, K.~(2013). `Realism, underdetermination and string theory dualities'. {\it Synthese}, 190 (3), pp.~471–489.\\
\\
M\o ller-Nielsen, T.~(2017). `Invariance, interpretation and motivation'. {\em Philosophy of Science} {\bf 84}, 1253-1264.\\
\\
Penrose, R.~(1982). `Quasi-local mass and angular momentum in general relativity'. {\it Proceedings of the Royal Society of London}, A 381, pp.~53-63.\\
\\
Pooley, O.~(2017). `Background Independence, Diffeomorphism Invariance, and the Meaning of Coordinates'. In: Lehmkuhl, D., Schiemann, G., Scholz, E.~(Ed.), {\it Towards a Theory of Spacetime Theories}, Birkh\"auser.\\
\\
Read, J.~(2016). `The Interpretation of String-Theoretic Dualities. {\it Foundations of Physics}, 46, pp.~209-235.\\
\\
Read, J.~and M\o ller-Nielsen, T.~(2018). `Motivating Dualities'. Forthcoming in {\it Synthese}.  http://philsci-archive.pitt.edu/14663\\
\\
Rickles, D.~(2017). `Dual theories: `same but different' or different but same'?' {\em Studies in History and Philosophy of Modern Physics}, 59, 62-67. doi:~10.1016/j.shpsb.2015.09.005.\\
\\
Szabados, L.~B.~(2009). `Quasi-Local Energy-Momentum and Angular Momentum in General Relativity'. {\it Living Reviews in Relativity}, 12, pp.~4.\\
\\   
van Fraassen, B. (1980). {\em The Scientific Image}. Oxford: Oxford University Press.\\
\\
van Fraassen, B.~C.~(2014). `One or Two Gentle Remarks about Hans Halvorson's Critique of the Semantic View'. {\it Philosophy of Science}, 81, pp.~276-283.\\
\\
Wang, M.-T.~(2015). `Four Lectures on Quasi-Local Mass'. arXiv:1510.02931 [math-ph].\\
\\
Weinberg, S.~(1995). {\it The Quantum Theory of Fields}, Volume I: Foundations. Cambridge: Cambridge University Press.

\end{document}